\documentclass[traditabstract]{aa}  
\usepackage{nameref}
\usepackage[switch]{lineno}
\usepackage{graphicx}
\usepackage{natbib}
\usepackage{amsmath} 

\newcommand{\kms}{$\rm{\,km \,s}^{-1}$}
\usepackage[english]{babel}
\usepackage[utf8]{inputenc}
\usepackage{amssymb}
\usepackage{xcolor}
\makeatletter

\newcommand{\Rmnum}[1]{\expandafter\@slowromancap\romannumeral #1@}
\makeatother
\usepackage[nottoc]{tocbibind}
\usepackage{txfonts}
\usepackage{verbatim} 
\usepackage[colorlinks=true,linkcolor=bluea,allcolors=blue]{hyperref}
\usepackage[font=normalsize,labelfont={bf}]{caption}
\usepackage{subfig}

\begin{document} 

\subtitle{
  Multi-phase characterization of AGN winds in 5 local type-2 quasars}
  \author{G. Speranza\inst{1,2}
  		\and C. Ramos Almeida\inst{1,2} 
		\and J. A. Acosta-Pulido\inst{1,2}
            \and A. Audibert\inst{1,2}
            \and L. R. Holden\inst{3}
            \and C. N. Tadhunter\inst{3}
            \and A. Lapi\inst{4}
            \and O. González-Martín\inst{5}
            \and M. Brusa\inst{6,7}
            \and I. E. López\inst{6,7}
            \and B. Musiimenta\inst{6,7}
            \and F. Shankar\inst{8}}

\institute{Instituto de Astrofísica de Canarias, Calle Vía Láctea, s/n, E-38205, La Laguna, Tenerife, Spain;\\
\email giovanna.speranza@iac.es 
\and Departamento de Astrofísica, Universidad de La Laguna, E-38206, La Laguna, Tenerife, Spain;
\and Department of Physics $\&$ Astronomy, University of Sheffield, S6 3TG Sheffield, UK;
\and SISSA, Via Bonomea 265, 34136 Trieste, Italy;
\and Instituto de Radioastronomía and Astrofísica (IRyA-UNAM), 3-72 (Xangari), 8701, Morelia, Mexico;
\and Dipartimento di Fisica e Astronomia "Augusto Righi", Università di Bologna, via Gobetti 93/2, 40129, Bologna, Italy;
\and INAF - Osservatorio di Astrofisica e Scienza dello Spazio di Bologna, via Gobetti 93/3, 40129, Bologna, Italy;
\and School of Physics and Astronomy, University of Southampton, Highfield, Southampton, SO17 1BJ, UK.
}

   \date{}

  \abstract {
  
  We present MEGARA (Multi-Espectrógrafo en GTC de Alta Resolución para Astronomía) Integral Field Unit (IFU) observations of five local type-2 quasars (QSO2s, z $\sim 0.1$) from the Quasar Feedback (QSOFEED) sample. These active galactic nuclei (AGN) have bolometric luminosities of 10$^{45.5-46}$ erg s$^{-1}$ and stellar masses of $\sim$10$^{11}$ M$_{\rm \sun}$. 
  The LR-V grating   of MEGARA allows us to explore the kinematics of the ionized gas through the [O~III]$\lambda$5007 $\AA$ emission line. The nuclear spectra of the five QSO2s, extracted in a circular aperture of $\sim 1.2 \arcsec$ ($\sim$ 2.2 kpc) in diameter,   matching the resolution of these seeing-limited observations, show signatures of high velocity winds in the form of broad (full width at half maximum;  1300$\leq$FWHM$\leq$2240 \kms) and blueshifted components.  
  We find that four out of the five QSO2s   present outflows that we can resolve with our seeing-limited data, and they have radii ranging from  3.1 to 12.6 kpc. In  the case of the two QSO2s with extended radio emission, we find that it is well-aligned with the outflows, 
  suggesting that  low-power jets might be compressing and accelerating the ionized gas in these radio-quiet QSO2s. In the four QSO2s with spatially resolved outflows, we measure ionized mass outflow rates of  3.3-6.5 M$_\odot$ yr$^{-1}$ when we use [S~II]-based densities, and of 0.7-1.6 M$_\odot$ yr$^{-1}$ when trans-auroral line-based densities are considered instead. We compare them with the corresponding molecular mass outflow rates (8 - 16 M$_\odot$ yr$^{-1}$), derived from CO(2-1) ALMA observations at 0.2\arcsec~resolution. 
  The cold molecular outflows carry more mass than their ionized counterparts. However, both phases show lower outflow mass rates than those expected from observational scaling relations  where uniform assumptions on the outflow properties were adopted. This might be indicating that the AGN luminosity is not the only driver of massive outflows and/or that these relations need to be re-scaled using accurate outflow properties (i.e., electron density and radius). 
  We do not find a significant impact of the outflows on the global star formation rates when considering the energy budget of the molecular and ionized outflows together. However, spatially resolved measurements of recent star formation in these targets are needed in order to fairly evaluate this, considering the dynamical timescales of the outflows, of 3-20 Myr for the ionized gas and 1-10 Myr for the molecular gas.

 }


\maketitle

\section{Introduction}
\label{introduction}

The evolution of massive galaxies is related to the accretion of matter onto super massive black holes (SMBHs). The SMBH is thought to self-regulate gas accretion by releasing energy, which has an impact on the host galaxy (e.g., \citealt{Fabian12}), as gas cannot cool down efficiently to form stars (\citealt{Dubois16}). This feedback from Active Galactic Nuclei (AGN) is required in cosmological simulations to regulate the growth of galaxies and match the observations (e.g., \citealt{Schaye15, Nelson18}). 
Outflows of gas are a manifestation of feedback, and they are capable of modifying the distribution of gas in the central hundreds of parsecs of galaxies (e.g., \citealt{Burillo21,Ramos22}), and altering the chemical enrichment of the circumgalactic medium up to hundreds of kiloparsecs from the SMBH (e.g., \citealt{Navarro21}).

The main feedback channels are the so-called ``quasar/radiative'' mode and the ``radio/kinetic'' mode. The first is supposed to act in AGN  having high accretion rates (\citealt{Fabian12}),  while the second one is  generally associated with powerful radio galaxies, with a lower accretion rate, where the gas is mechanically compressed and accelerated by jets (\citealt{McNamara07}). However, this distinction of the two main feedback mechanisms is too simplistic, and both modes can act simultaneously. In fact, in radio-quiet type-2 quasars (QSO2s, L$_{\text{[O~III]}}>10^{8.5}\,\text{L}_{\odot}$; \citealt{Reyes08}) low-power jets have been proposed to accelerate outflows (e.g., \citealt{Jarvis19, Girdhar22, 2023arXiv231103453G, Speranza22, Venturi23}), even though the quasar mode should dominate due to their high accretion rates.  In this context, QSO2s are good laboratories to test whether compact low-power jets  can efficiently drive gas outflows (e.g., \citealt{Audibert23}). This can be explored by inspecting the spatial distribution and geometry of the radio emission with respect to the outflowing gas and comparing their energetics.

To detect signatures of  ionized outflows, it is common to characterize the [O~III]$\lambda 5007 \AA$ emission line that, being a forbidden transition, never has a broad component from the broad line region (BLR) and therefore is a good tracer of the kinematics of the narrow line region (NLR), from the central parsecs to kiloparsec scales (e.g., \citealt{Osterbrock89}). Any broad [O~III]$\lambda 5007 \AA$ profiles having prominent and asymmetric wings, can be associated with turbulent and fast outflowing gas that is not ordinarily rotating, as claimed in early (e.g., \citealt{Heckman81, Feldman82, Whittle85}) and recent studies (e.g., \citealt{Cresci15, Brusa15, Carniani15, Venturi18, Speranza21, Peralta23}). These ionized outflows are commonly observed in nearby QSO2s (e.g., \citealt{Liu13436,  Mullaney13, Harrison14, Villar14, Fischer18}) where the contamination from the continuum and the BLR is naturally masked by nuclear dust (\citealt{Ramos17a}), although large-scale obscuration might be important as well.

However, the ionized emission represents only a single phase of the gas, and to fully characterize the impact of outflows a multi-phase characterization is needed (\citealt{Cicone18}). From observational scaling relations, the cold molecular outflows are expected to  drive more massive outflows than the ionized ones (e.g., \citealt{Fiore17}) and the outflow properties, including densities and velocities, have been observed to be different depending on the gas phase (e.g., \citealt{Finlez18, Herrera19, Fluetsch21, Garc21}). However, the majority of works are based on a single outflow phase, providing a partial view of AGN feedback in which the interplay between the different phases of the gas is still  to be unveiled.

Here we present new integral field Unit (IFU)
data  of five QSO2s observed with the MEGARA instrument of the 10.4 m Gran Telescopio Canarias (GTC). This instrument provides optical spectra in which, at the redshift of the targets ( z$\sim$0.1), the [O~III]$\lambda 5007 \AA$ emission line is detected at high spectral resolution (R$\sim$5850 using the low-resolution grating). These QSO2s have already been observed with the Atacama Large Millimeter/submillimeter Array (ALMA)  at 0.2$\arcsec$ resolution and their cold molecular gas kinematics have been characterized by \citet{Ramos22} using the CO(2-1) emission line. These authors reported the presence of cold molecular outflows for the five QSO2s, four of them spatially resolved  with projected radii of $\sim$0.2-0.7$\arcsec$ ($\sim$0.4-1.3 kpc). Here we aim to study the ionized kinematics of the QSO2s and compare them with their cold molecular gas counterparts. The paper is organized as follows: in Section~\ref{sample} we present the main properties of the sample. Section~\ref{observations} includes the details on the observations and data reduction. In Section~\ref{analysis} we show the methodology applied to analyze the nuclear and extended emission of the QSO2s and to measure their outflow properties, and we describe the results. In Section~\ref{discussion} we discuss the implication of our findings, which we summarize in Section~\ref{conclusion}. Throughout this work we assume the following cosmology: H$_0$ = 70.0 \kms Mpc$^{-1}$, $\Omega_{\text{M}}$ = 0.3 and $\Lambda$ = 0.7.

\section{Sample selection and characteristics}
\label{sample}

\begin{table*}
  \caption{Properties of the five QSO2s.} 
  \begin{center}
\begin{tabular}{c c c c c c c c c}
\hline
\hline

SDSS ID    & z   &  Scale  &  log L$_{\text{[O~III]}}$  & log L$_{\text{bol}}$ &  log L$_{1.4\text{GHz}}$ &  log M$_*$ & SFR & Galaxy \\
 & SDSS & [kpc~arcsec$^{-1}$] &  [erg s$^{-1}$] & [erg s$^{-1}$] & [W Hz$^{-1}$] & [M$_{\odot}$] & [M$_{\odot}$yr$^{-1}$] & morphology
\\
\hline

 J101043.36+061201.4 & 0.0977 & 1.807 &  42.88   &  45.55 & 24.37 & 10.99$\pm$0.20 & 30 & Interacting ETG
 \\
 
 J110012.39+084616.3 &  0.1004  & 1.851   & 43.18 &  45.85 & 24.18 & 11.02$\pm$0.22 & 34 & Barred spiral
 \\
 
J135646.10+102609.0 & 0.1232 & 2.213     & 42.87 & 45.54 & 24.36 & 11.27$\pm$0.19 & 69 & Merging ETG
 \\

J143029.88+133912.0 & 0.0851 & 1.597 &  43.16   &  45.83   & 23.67 & 11.15$\pm$0.11 & 12 & Post-merger ETG
 \\

J150904.22+043441.8 & 0.1115 & 2.028    & 43.37  & 46.04 & 23.81 & 10.94$\pm$0.31 & 34 & Barred spiral
 \\

\hline
\end{tabular}
\end{center}
Notes: The bolometric luminosities were calculated by applying the correction factor of 474 from \citet{Lamastra09} to the extinction-corrected [O~III] luminosities from \citet{Kong18}. 
The 1.4 GHz luminosities, stellar masses, SFRs, and galaxy morphologies are from \citet{Ramos22}. 
\label{tab:sample} 
\end{table*}

\begin{figure*}
\centering
{\par\includegraphics[width=1.\textwidth]{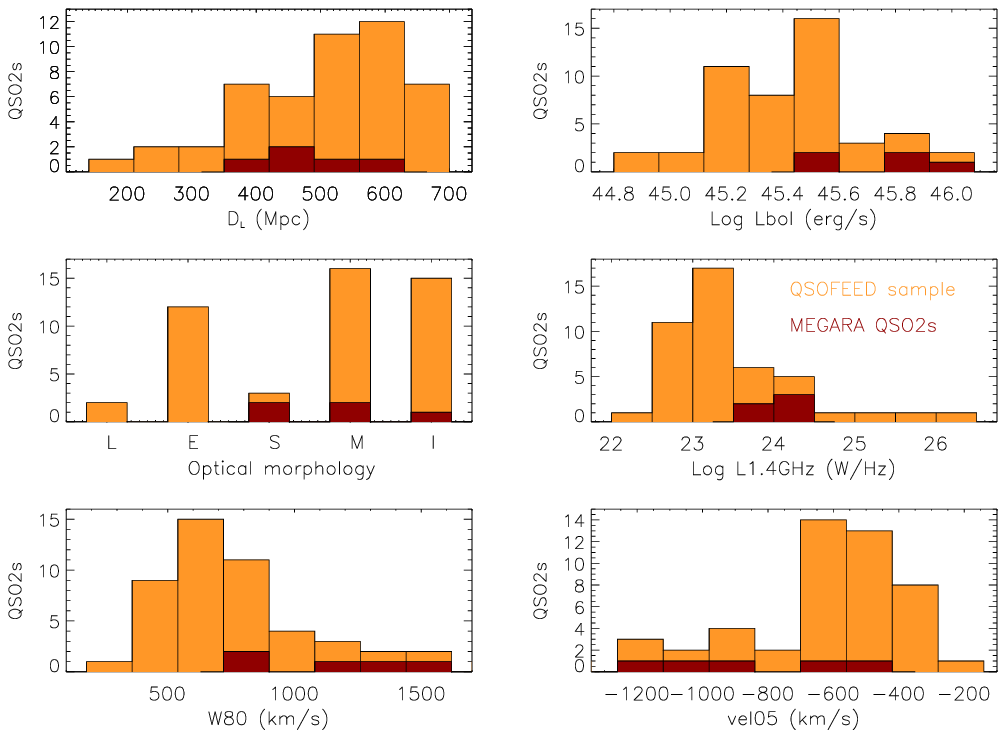}
}
\caption {  Orange and red histograms correspond to all QSO2s in the QSOFEED sample and to the 5 QSO2s studied here. From left to right and from top to bottom: luminosity distance, AGN luminosity, optical morphology (L=lenticular, E=elliptical, S=spiral, M=merger, I=interacting), integrated radio luminosity from FIRST 1.4 GHz data, W$_{80}$, and vel$_{05}$ of the [O~III] line profile measured from SDSS optical spectra of the targets ({\color {blue} Bessiere et al. in prep.}).   SDSS~J134733.36+121724.3 was excluded from the W$_{80}$ and  vel$_{05}$ histograms because of its extreme kinematics (W$_{80}=2500$\kms~and vel$_{05}=-2500$ \kms).}
\label{fig:sample}
\end{figure*}

The five QSO2s studied here are part of the Quasar Feedback (\href{http://research.iac.es/galeria/cra/qsofeed/}{QSOFEED}) sample \citep{Ramos22,Pierce23},  selected from the \citet{Reyes08} compilation of narrow-line AGN to have L$_{\text{[O~III]}} > 10^{8.5}$L$_{\odot}$ and redshift of z<0.14.   These selection criteria lead to a final number of 48 QSO2s with   an average stellar mass  and standard deviation of M$_*$ = 10$^{11.1\pm0.2}$ M$_\odot$, obtained from the 2MASS Extended Source Catalogue (XSC) K-band magnitudes (\citealt{Ramos22}). Our subset of five QSO2s have redshifts 0.08 $\leq$ z $\leq$ 0.12 and their main properties are listed in Table~\ref{tab:sample}.  We selected these 5 QSO2s because they were detected in the CO(2-1) ALMA observations reported in \citet{Ramos22}. Another two QSO2s, hosted in red early-type galaxies, were observed with ALMA at the same angular resolution, but they were not detected in CO(2-1) and therefore have molecular gas masses <7$\times$10$^{8}$ M$_{\sun}$. Thus, the 5 QSO2s studied here are representative of the gas-rich QSO2s in the QSOFEED sample, as they have total molecular gas masses of 4-18$\times$10$^{9}$ M$_{\sun}$. Similar gas masses, of $\sim$1$\times$10$^{10}$ M$_{\sun}$, were reported by \citet{2020MNRAS.498.1560J}, based on APEX CO(2-1) observations a sample of QSO2s at z<0.2 hosting kpc-scale ionized outflows and jets. 

  In Fig.~\ref{fig:sample} we show how our subset of 5 QSO2s compares with the whole QSOFEED sample in terms of luminosity distance (D$_{\text{L}}$), bolometric luminosity, optical morphology, 1.4 GHz luminosity, and [O~III] kinematics (W$_{80}$ and vel$_{05}$; i.e., velocity dispersion and outflow velocity).   
 The luminosity distances of the 5 QSO2s are intermediate compared with those of the QSOFEED sample, and the bolometric luminosities, derived from the extinction-corrected [O III] luminosities from \citet{Kong18} by applying the correction factor of 474 from \citet{Lamastra09}, are among the highest (log L$_\text{bol} = 45.5-46.0$ erg~s$^{-1}$; see top panels of Fig. \ref{fig:sample}). The 5 QSO2s have diverse optical morphologies, with three of them showing signatures of galaxy interactions and mergers. In this context, they are representative of the QSOFEED sample, which includes at least 65\% of interacting and merging galaxies 
(\citealt{Pierce23}). As can be seen for the middle left panel of Fig. \ref{fig:sample}, we lack undisturbed lenticular and elliptical galaxies. 

Our subset of QSO2s have radio luminosities that are intermediate (log L$_{1.4\text{GHz}}$=[23.7,24.4] W~Hz$^{-1}$) within the range probed by the QSOFEED sample, although the majority of the QSO2s have radio luminosities in the range [22.5,23.5] W~Hz$^{-1}$. It is noteworthy that four of the QSO2s in the QSOFEED sample 
qualify as radio-loud objects, 
having log L$_{1.4\text{GHz}}>$24.5 W~Hz$^{-1}$ (see the middle right panel of Fig. \ref{fig:sample}) and L$_{1.4\text{GHz}}$/L$_{\text{[O~III]}}$ ratios above the \citet{Xu99} division. The other 44 QSO2s in the QSOFEED sample are radio-quiet, including the 5 QSO2s studied here. The relatively high radio luminosities of the latter put them above the radio-FIR correlation of star-forming galaxies (\citealt{Bell03}). According to \citet{Jarvis19}, in the case of the four QSO2s that we have in common (J1010, J1100, J1356, and J1430), less than 7\% of their radio emission is associated with star formation, with non-thermal AGN emission being responsible for the rest. 
\citet{Ramos22} estimated the SFRs of the QSO2s from their FIR luminosities, following \citet{Kennicutt98}, corrected to a Chabrier initial mass function (i.e., dividing by a factor of 1.59; \citealt{Chabrier03}). These SFRs are supposed to be less affected by the AGN contribution than estimations from mid-infrared or radio, and therefore we adopt these values here. The SFRs range between 12 and 69 M$_{\odot}$~yr$^{-1}$,  with uncertainties of $\sim$0.3 dex, and they are comparable with the values reported by \citet{Jarvis19} for J1010, J1100, J1356, and J1430 (8 - 84 M$_{\odot}$~yr$^{-1}$), measured from the 
IR luminosity due to star formation, derived from spectral energy distribution (SED) fitting.

Finally, the bottom panels of Fig.~\ref{fig:sample} provide a comparison of the [O~III] kinematics of the QSOFEED and MEGARA samples. To ensure a fair comparison, we fitted the [O III] lines detected in SDSS spectra of the 48 objects in the QSOFEED sample, including the 5 MEGARA QSO2s, using the same methodology presented in Section~\ref{extended}, where we define W$_{80}$ and vel$_{05}$.  The MEGARA QSO2s are at the high-velocity dispersion and high-velocity end of the two histograms, particularly in the case of W$_{80}$. Therefore, the five QSO2s studied here are representative of the QSO2s having the highest AGN luminosities and the most extreme ionized gas kinematics in the QSOFEED sample. 


 The ionized outflows of the five QSO2s have been studied using data from different facilities operating in the optical or near-infrared (NIR): the Gemini-South GMOS IFU (J1010, J1100, J1356, and J1430; \citealt{Harrison14}), the Low-Dispersion Survey Spectrograph (J1356; \citealt{Greene12}), HST/STIS (J1100; \citealt{Fischer18}), VLT/SINFONI (J1430; \citealt{Ramos17}), and GTC/EMIR (J1509; \citealt{Ramos19}). 
In the case of the last two datasets, the NIR spectra enable the simultaneous characterization of the ionized gas emission from the Pa$\alpha$ and [Si~VI]$\lambda1.963 \mu$m lines and the warm molecular gas from the H$_2$ lines.   However, only J1509 shows a clearly detected warm molecular outflow. 
 Finally, \citet{Ramos22} reported signatures of cold molecular outflows in all the QSO2s, four of them spatially resolved (J1010, J1100, J1356, and J1430).
%

\section{MEGARA observations and data reduction}
\label{observations}

\begin{table}
  \caption{Summary of the GTC/MEGARA observations.} 
  \begin{center}
\begin{tabular}{c c c c c}
\hline
ID    & Exp.   &  Seeing  & Obs. date  & Airmass  \\
\hline
 J1010 & 3$\times$403 s & 1.1$\arcsec$     & ~~2021 March 16    & 1.1
 \\
 
 J1100 &  3$\times$247 s  & 1.0$\arcsec$   & 2021 April 02 &  1.2
 \\
 
J1356 & 3$\times$488 s & 1.1$\arcsec$     & ~~2021 March 18  &  1.1
 \\

J1430 & 3$\times$120 s & 1.1$\arcsec$    &  ~~2021 March 18   & 1.2
 \\

J1509 & 3$\times$590 s & 1.1$\arcsec$    & ~~2021 March 22  & 1.2
 \\

\hline
\end{tabular}
\end{center}
Column description: (1) object ID; (2) exposure time; (3) seeing measured from the FWHM of the standard star observed just before each QSO2; (4) day of observation; (5) airmass.
\label{tab:obs} 
\end{table}

The sample was observed during four nights on 2021, between March and April (see Table~\ref{tab:obs}; program ID: GTC62-20B, PI: C. Ramos Almeida), using the optical IFU mode of MEGARA that is installed on the Nasmyth A focus of the GTC. The IFU Field-of-view (FOV) of MEGARA is 12.5$\arcsec$ $\times$ 11.3$\arcsec$ ($\sim$ 23.0 $\times$ 20.8 kpc$^2$ at z=0.1) and it is composed by 567 hexagonal fibres, each one circumscribed in a circle of 0.62$\arcsec$ in diameter. Additional 56 fibres, divided in 8 minibundles, are located at 1.75 - 2 arcmin from the center of the FOV to simultaneously measure the sky background during observations. We made use of one of the low resolution volume phase holographic gratings (VPH570-LR; hereafter LR-V), which has a wavelength coverage of 5144-6168 $\AA$ with a nominal spectral resolution of R=5850 ($\sim$ 51 \kms) at the central wavelength of $\lambda_c = $5687.63 $\AA$. Using our own dataset, we measured a median FWHM = 0.9 $\AA$ from the ArXe lamps that corresponds to a spectral resolution of $\sim$ 45 \kms. We use the corresponding values measured for each source from the lamps as our instrumental broadening. 

The integration times for each target are reported in Table~\ref{tab:obs}. We took three exposures of a maximum of 590 s of integration to reduce the impact of cosmic rays. During the entire observation run the conditions were clear  (i.e., less than 10$\%$ of the sky covered by cirrus), except for a photometric night when J1010 was observed. We calculated the seeing full width at half maximum (FWHM) using the photometric standard stars observed just before  each target. The seeing was FWHM = 1.1$\arcsec$ ($\sim$2 kpc) for the whole sample, except for J1100, which has a slightly lower value of 1.0$\arcsec$ (see Table~\ref{tab:obs}). 

The data reduction was performed using the MEGARA pipeline (\citealt{Pascual20, Pascual21}). This pipeline performs bias subtraction, bad-pixels masking, tracing fibres, wavelength calibration (including versions corrected and uncorrected of barycentric velocity), flatfield correction, flux calibration, sky subtraction, and cube reconstruction. As recommended by the pipeline developers, the final data cubes were produced with spaxels of 0.3$\arcsec$ in size.

\section{Methodology and results}  \label{analysis}
\label{results}

\subsection{The nuclear [O~III] emission} \label{nucleus}

\begin{figure*}[h!]
\centering
\includegraphics[width=0.52\textwidth]{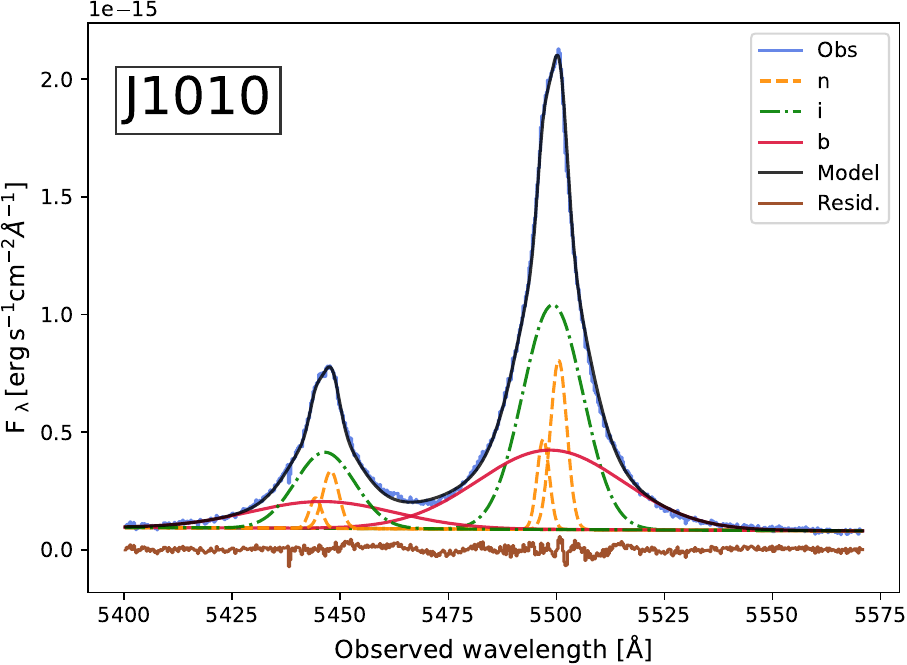}
\includegraphics[width=0.475\textwidth]{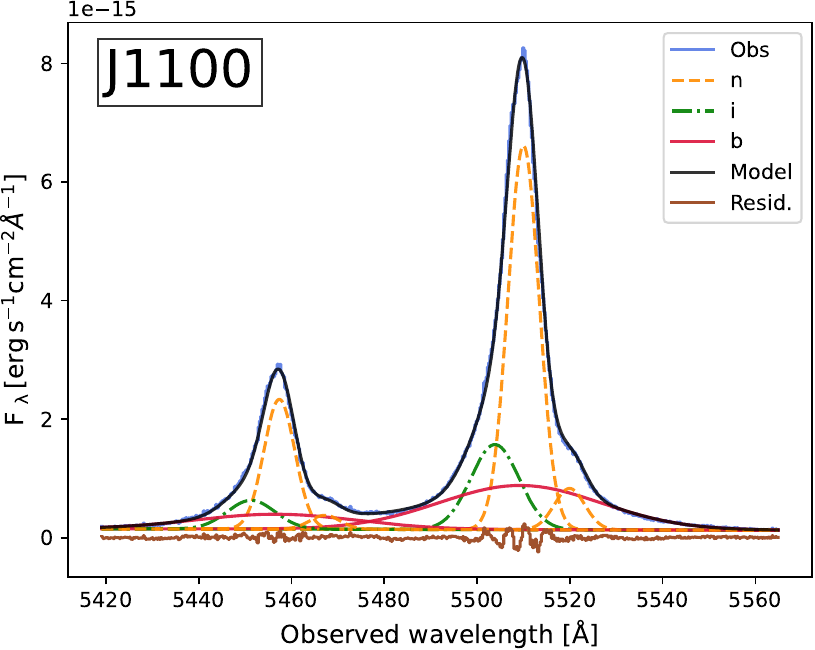}
\includegraphics[width=0.49\textwidth]{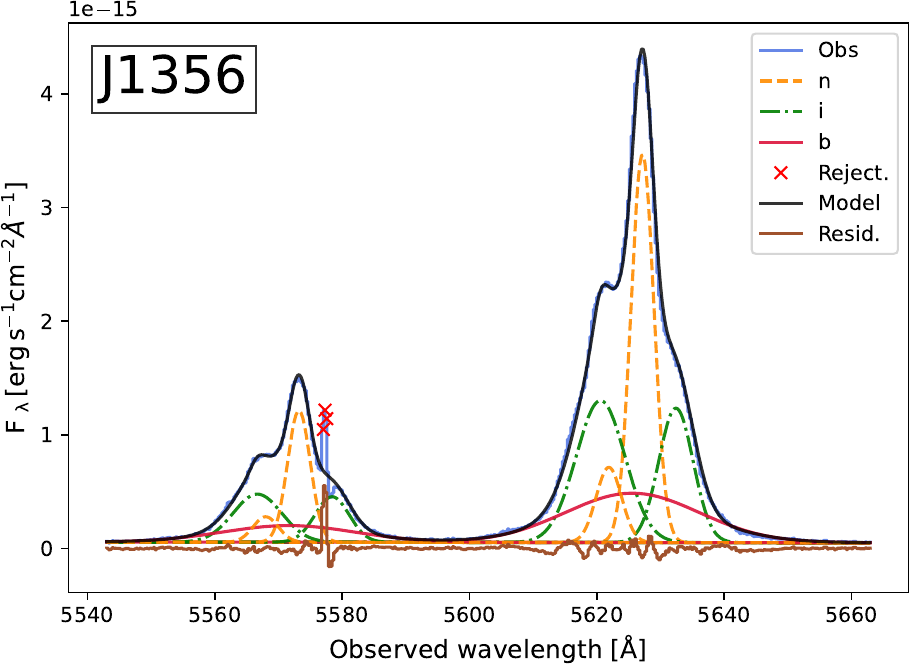}
\includegraphics[width=0.49\textwidth]{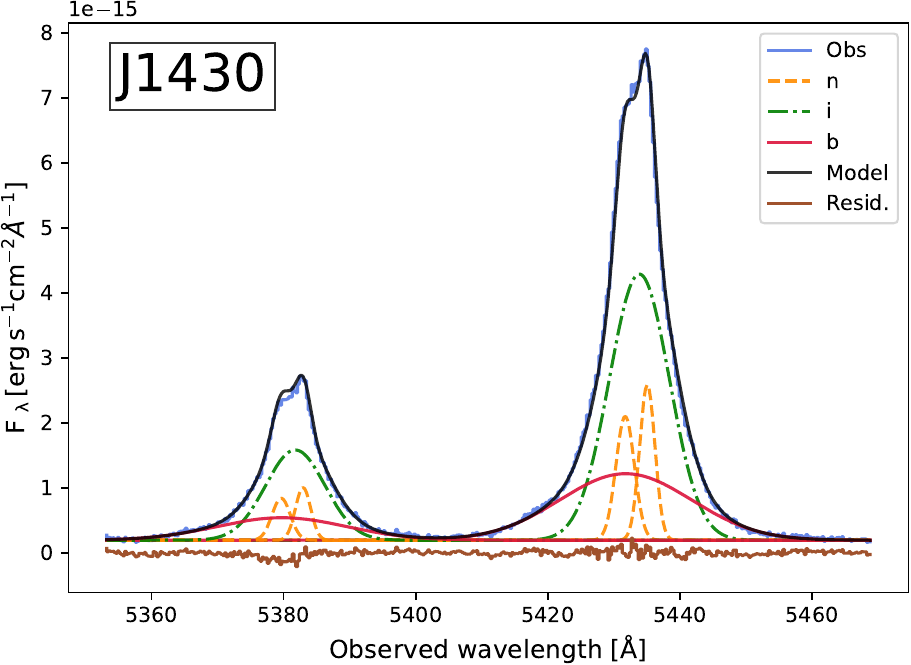}
\includegraphics[width=0.49\textwidth]{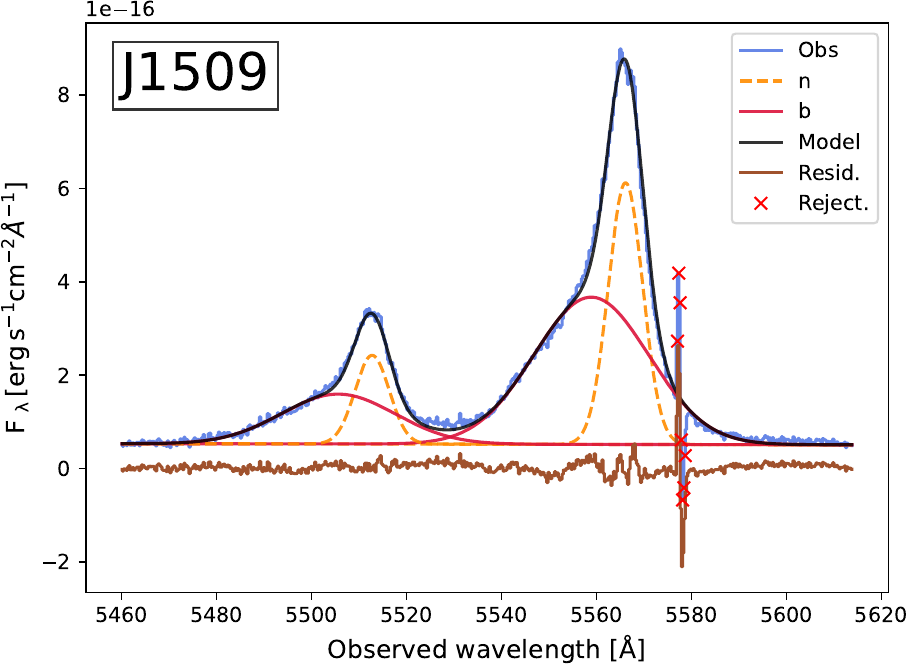}
\caption {Fits of the [O~III] doublets observed in the nuclear spectra extracted with an aperture of 1.2\arcsec~in diameter. The observed spectra are shown in blue, the model including all the components in black, and the residuals from the fit in brown. Same colors correspond to the same physical components: narrow (n; orange dashed), intermediate (i; green dot-dashed), and broad (b; red solid). 
Noisy pixels not considered in the fits are marked with red crosses.  
}
\label{fig:nucfit}
\end{figure*}

In order to study the ionized gas kinematics we firstly extracted the nuclear spectra of the five QSO2s in a circular aperture of 1.2\arcsec~($\sim$2.2 kpc)  in diameter. We did this to match the angular resolution set by the seeing size (see Table~\ref{tab:obs}). Thanks to the LR-V wavelength coverage of MEGARA,  H$\beta$ and the [O~III]$\lambda\lambda$4959,5007 \AA~doublet are detected at the redshift of our sample. Considering that the [O~III]$\lambda$5007 \AA~ line is the most prominent, we extracted the nuclear spectra centering them on its peak, that always coincides with the continuum peak except in J1356, where it is off-setted by one spaxel.

We characterized the [O~III] doublet by simultaneously fitting multiple Gaussians after subtracting the contribution of the continuum. We selected the continuum in the blue and red side adjacent to the [O~III] doublet, so as to exclude any flux contribution from the emission lines, and we performed a linear fit of the selected regions to then subtract it. The width of each of these bands and the distance from the line peak depends on the broadness of the doublet and on the proximity of the H$\beta$ emission line. The multiple Gaussians were fitted using the same in-house Python program described in Section 3.1 of \citet{Speranza22}, that is based on the modelling module of \textit{Astropy} (\citealt{Astropy13,Astropy18}). For this work, we select the best fit as the model with the minimum number of Gaussian components for which the reduced-$\chi^2$ improves more than 10$\%$ by adding the last component, as in \citet{Bessiere22}.  We fit two Gaussians in the case of J1509,  four in J1010, J1100, and J1430, and five in J1356. Data points having a standard deviation higher than 10 times the standard deviation of the residuals were removed (see the fits of J1356 and J1509 in Fig.~\ref{fig:nucfit}). This threshold 
was selected by performing a preliminary fit of the [O~III] doublet from which we evaluated the points clearly detached from the line profiles, which most likely correspond to hot pixels. 

In all the five QSO2s the best fits include at least two Gaussians: a narrow component with FWHM between 150 and 460 \kms, and a broad (FWHM$\ge$1300 \kms) and   component blueshifted from the systemic velocity set by the narrow component(s) (velocity shift, v$_s$, between -40 and -430 \kms). In four of the QSO2s we also detect intermediate components with FWHM = 340 - 890 \kms, most of them blueshifted (between -20 and -370 \kms) but also redshifted (26 and 280 \kms~in J1430 and J1356, respectively). We note that J1356 has an intermediate component that is narrower than the narrow components of J1100 and J1509 (see Table \ref{tab:nuc}). Here we define the intermediate components as those being at least 100 \kms~broader in FWHM than the narrowest component fitted to the same line, but still significantly narrower than the broadest component. All the fits are shown in Fig.~\ref{fig:nucfit} and the parameters are listed in Table~\ref{tab:nuc}.

In general, we associate the narrow components with gas in the NLR. In all cases but J1509 we resolve two narrow components that might correspond to the approaching and receding sides of a rotating gas distribution (see Section \ref{extended}), although some of the QSO2s show complex kinematics (J1356 and J1430),  suggesting that ongoing mergers and/or the presence of jet-ISM interactions might be altering or preventing gas rotation. 
On the other hand, the broad components correspond to turbulent outflowing  gas, since they have FHWM$\ge$1300\kms~and they are blueshifted with respect to the systemic velocity. We note that these blueshifts are small in the case of the nuclear spectra because we are looking at the integrated spectrum of the central 1.2 arcsec of the galaxies, but taking advantage of the spatial information provided by the IFU, we measure velocities of $\ge$800 \kms~for all the QSO2s (see Appendix \ref{AppendixA}). Therefore, we detect the previously reported ionized outflows in the 5 QSO2s using this high spectral resolution dataset. 
The case of the intermediate components is more ambiguous since in some cases their FWHMs and v$_s$ are consistent with those of the NLR (e.g., J1356), and in others with outflowing gas (e.g., J1100). For this reason in Section \ref{extended} we use a non-parametric method to measure the outflow properties, rather a parametric one. By doing this we do not have to ascribe any physical meaning to each kinematic components.

In Table \ref{tab:nuc} we report the results from the [O III] fits   performed using Gaussians. In particular, we list the FWHM, the velocity shift from the narrow component (v$_s$), and the integrated flux. In the case of double peaked profiles (i.e., two narrow components detected with comparable fluxes), the systemic velocity corresponds to the flux weighted value between the two peaks (J1010 and J1430). Instead, when a dominant narrow component is fitted, it is used to obtain the systemic velocity (J1100, J1356, and J1509). 
In Table \ref{tab:nuc} and throughout the text, we refer only to the parameters of [O III]$\lambda$5007 \AA,  since it shares the same number of components, each with identical FWHM and v$_s$, and an intensity ratio of 2.98  with [O~III]$\lambda$4959 \AA~(\citealt{Storey00}), as they are part of the same doublet. 
The uncertainties of the parameters were measured by means of a Monte Carlo simulation. We simulate mock spectra by varying the flux of each spectral element of the emission line profile by adding random values extracted from a normal distribution with an amplitude given by the noise of the pixels. The uncertainties of the parameters are then computed as 1$\sigma$ of each parameter distribution obtained from 100 mock spectra.

\begin{table*}
  \caption{Main properties of the [O~III] fits from the nuclear spectra of the QSO2s.} 
  \begin{center}
\begin{tabular}{c c c c c}
\hline
ID    & Line   &  FWHM  & v$_s$  & Flux $\times10^{-15}$  \\
      &  $ [\text{O~III}]     $    &  [km s$^{-1}$]   & [km s$^{-1}$]  &  [erg cm$^{-2}$s$^{-1}$]\\
\hline
 J1010 &   (n) & $~~~~238\pm10$     & $~~~~~61\pm3$    & $~~3.45\pm0.23$
 \\
 $\lambda_c$ = 5499.5 \AA &      (n) & $~~175\pm6$     & $-153\pm6$  & $~~1.35\pm0.08$\\
&       (i) & $~~~~892\pm30$     & $~~-21\pm2$  & 
$16.70\pm0.71$\\
&         (b) & $~~2164\pm80$     & $~~-55\pm7$  & $14.19\pm1.73$\\
\hline
 J1100 &   (n) & $429\pm1$     & $~~~~~~~0\pm1$   & $54.62\pm0.22$ 
 \\
 $\lambda_c$ = 5510.0 \AA  &       (n) & $420$     & $~~~594\pm2$    & $~~5.77\pm0.07$
\\
  &    (i) & $664\pm5$     & $-366\pm3$   & $18.66\pm0.31$
 \\
&      (b) & $2241\pm13$     & $~~-40\pm2$   & $32.74\pm0.83$
\\

\hline
J1356 &  (n) & $226\pm1$     & $~~~~~~~~0\pm1$   &  $15.46\pm0.10$
 \\
$\lambda_c$ = 5627.3 \AA &    (n) & $250\pm4$     & $~-281\pm3$   & $~~3.31\pm0.05$
 \\
&    (i) & $338\pm5$     & $~~~~280\pm2$   & $~~8.00\pm0.14$
 \\
&   (i) & $483\pm5$     & $~-351\pm2$   & $12.03\pm0.16$
 \\
&       (b) & $1325\pm25$     & $~~~-85\pm4$  & $11.50\pm0.47$
\\
\hline
J1430 &    (n) & $151\pm5$     & $~~~~97\pm2$   & $~~6.97\pm0.24$
 \\
$\lambda_c$ = 5433.4 \AA &         (n) & $181\pm6$     & $~~-102\pm3$   & $~~6.70\pm0.22$
\\
&   (i) & $590\pm7$     & $~~~~26\pm2$   &  $46.59\pm1.06$
 \\
&       (b) & $1296\pm23$     & $~~-98\pm8$  & $25.58\pm2.06$
\\
\hline
J1509 &  (n) & $~~458\pm4$     & $~~~~~~~0\pm1$  & $~~5.10\pm0.06$
 \\
$\lambda_c$ = 5558.9 \AA &        (b) & $1536\pm7$     & $-431\pm1$  & $~~9.57\pm0.11$
\\
\hline
\end{tabular}
\end{center}
Column description: (1) object ID followed by the observed wavelength of [O~III] ($\lambda_c$); (2) kinematic component: narrow (n), intermediate (i) or broad (b);  (3) FWHM corrected from instrumental broadening; (4) velocity shifts (v$_s$) measured from $\lambda_c$; (5) integrated line flux. Measurements without errors correspond to parameters that have been fixed.
\label{tab:nuc} 
\end{table*}


\subsection{A spatially resolved analysis}
\label{PSF} 

\begin{figure}[h!]
\centering
    \subfloat{{\includegraphics[width=4cm]{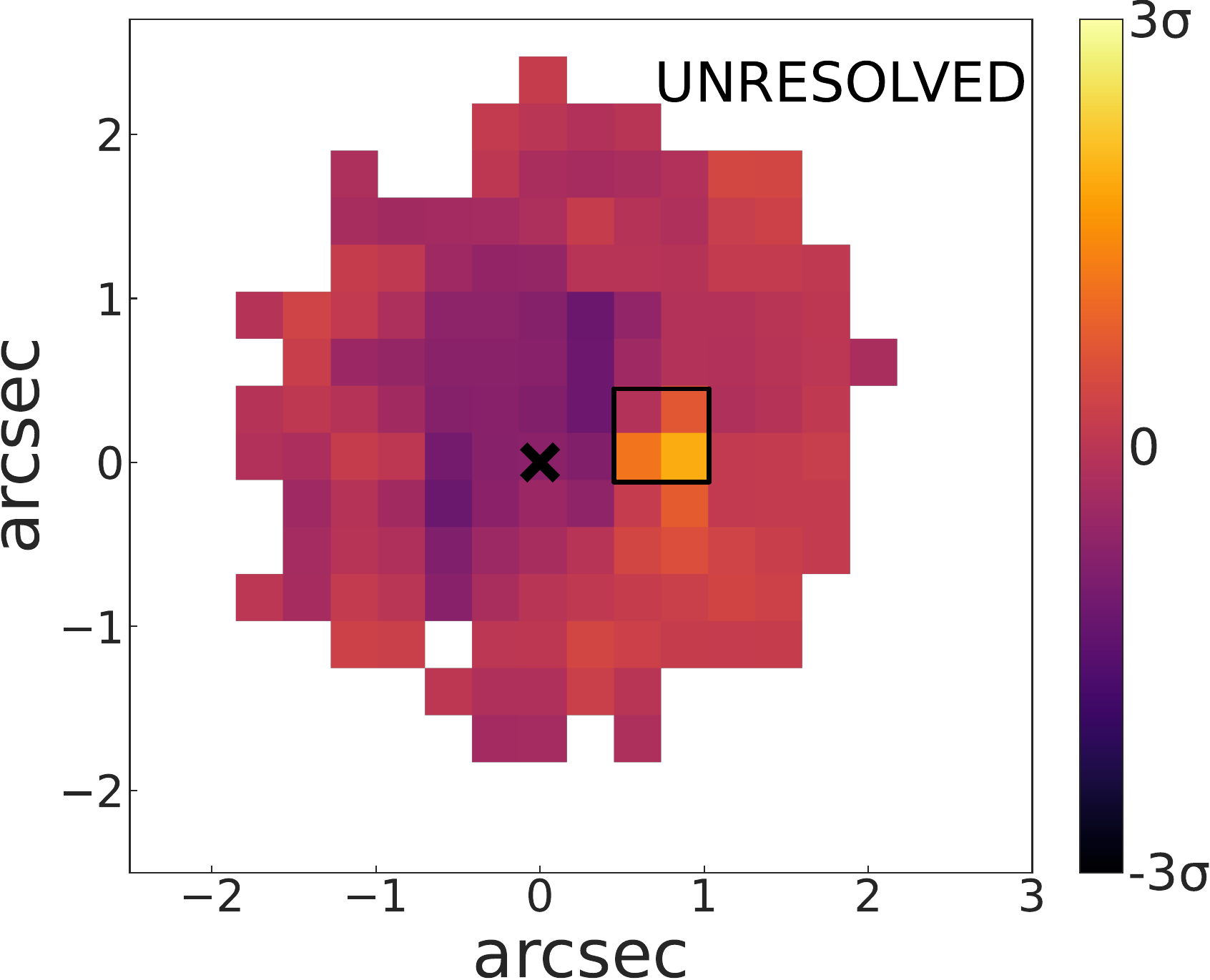} }}%
    \qquad 
    \subfloat{{\includegraphics[width=4.12cm]{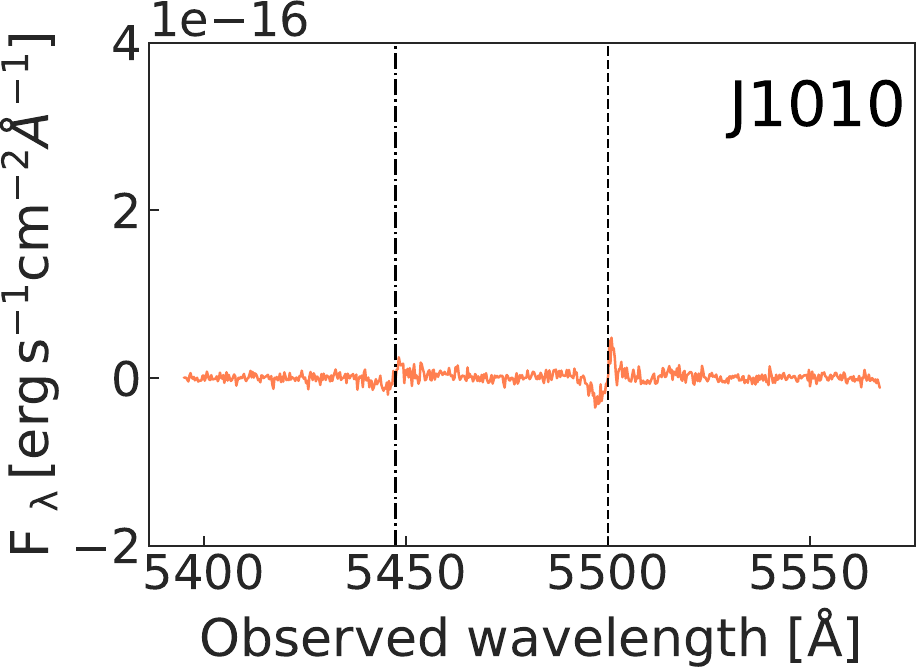} }}
    \qquad
    \subfloat{{\includegraphics[width=4.cm]{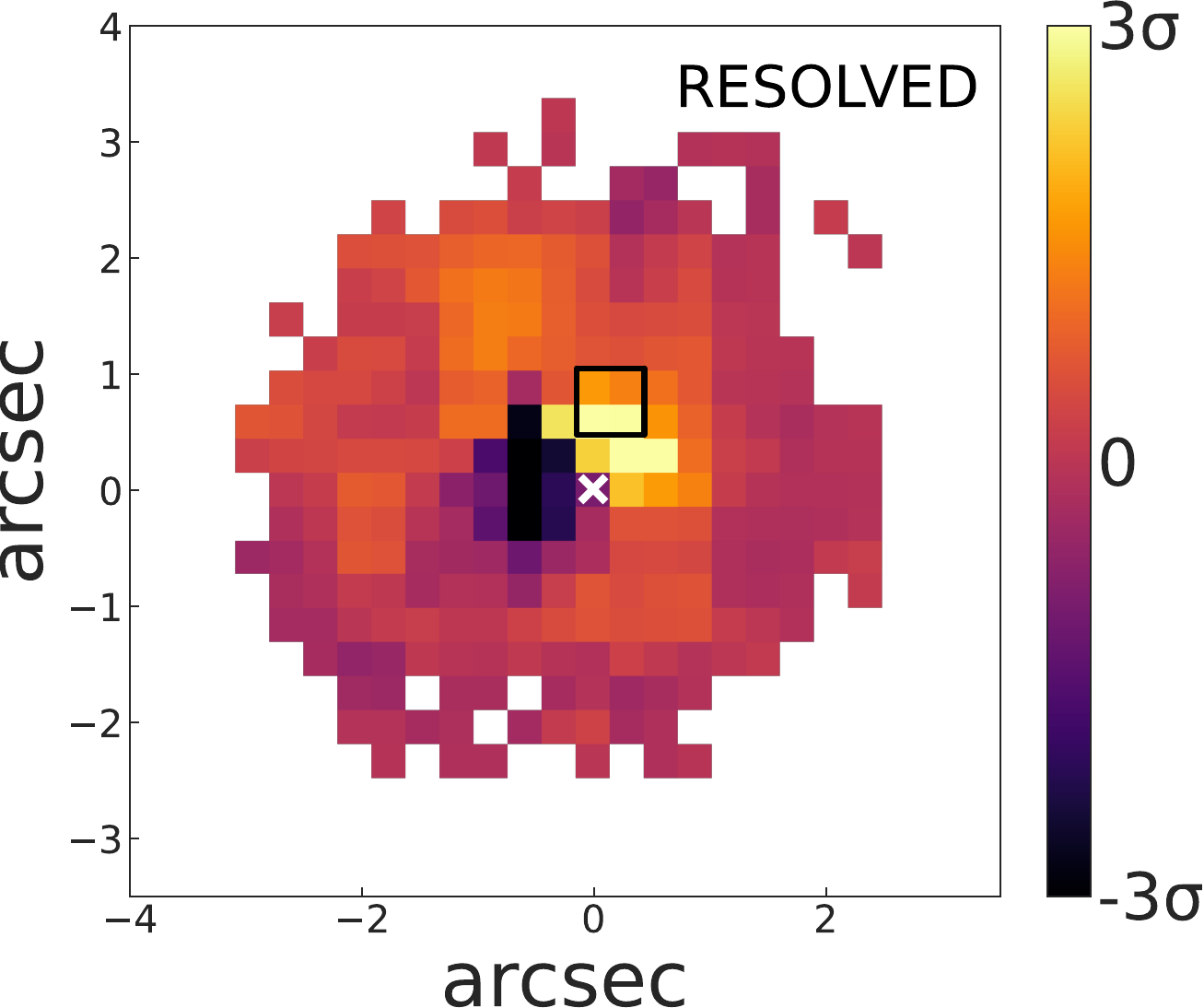} }}
    \qquad
    \subfloat{{\includegraphics[width=4.12cm]{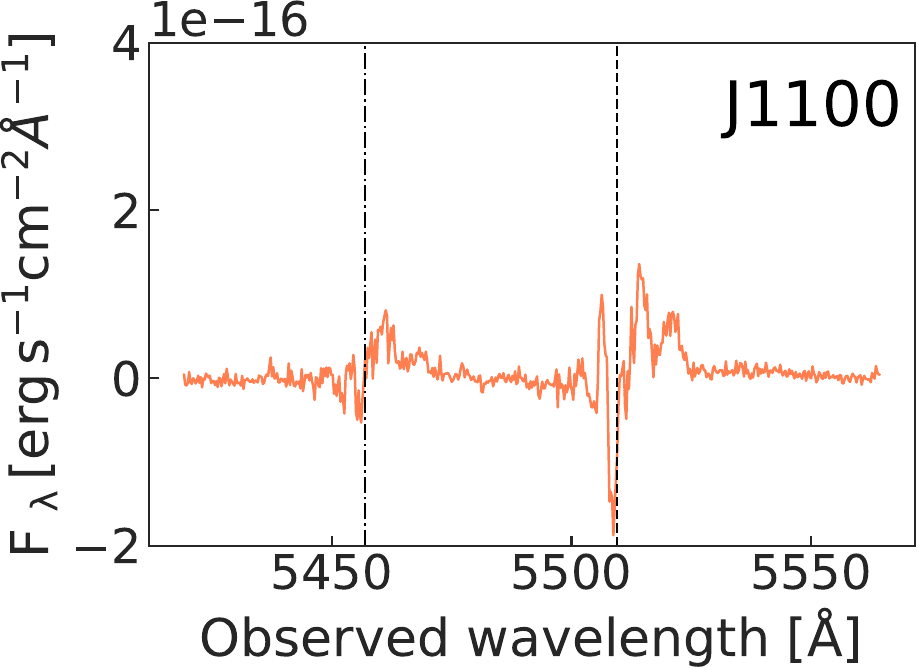} }}
    \qquad
    \subfloat{{\includegraphics[width=4.cm]{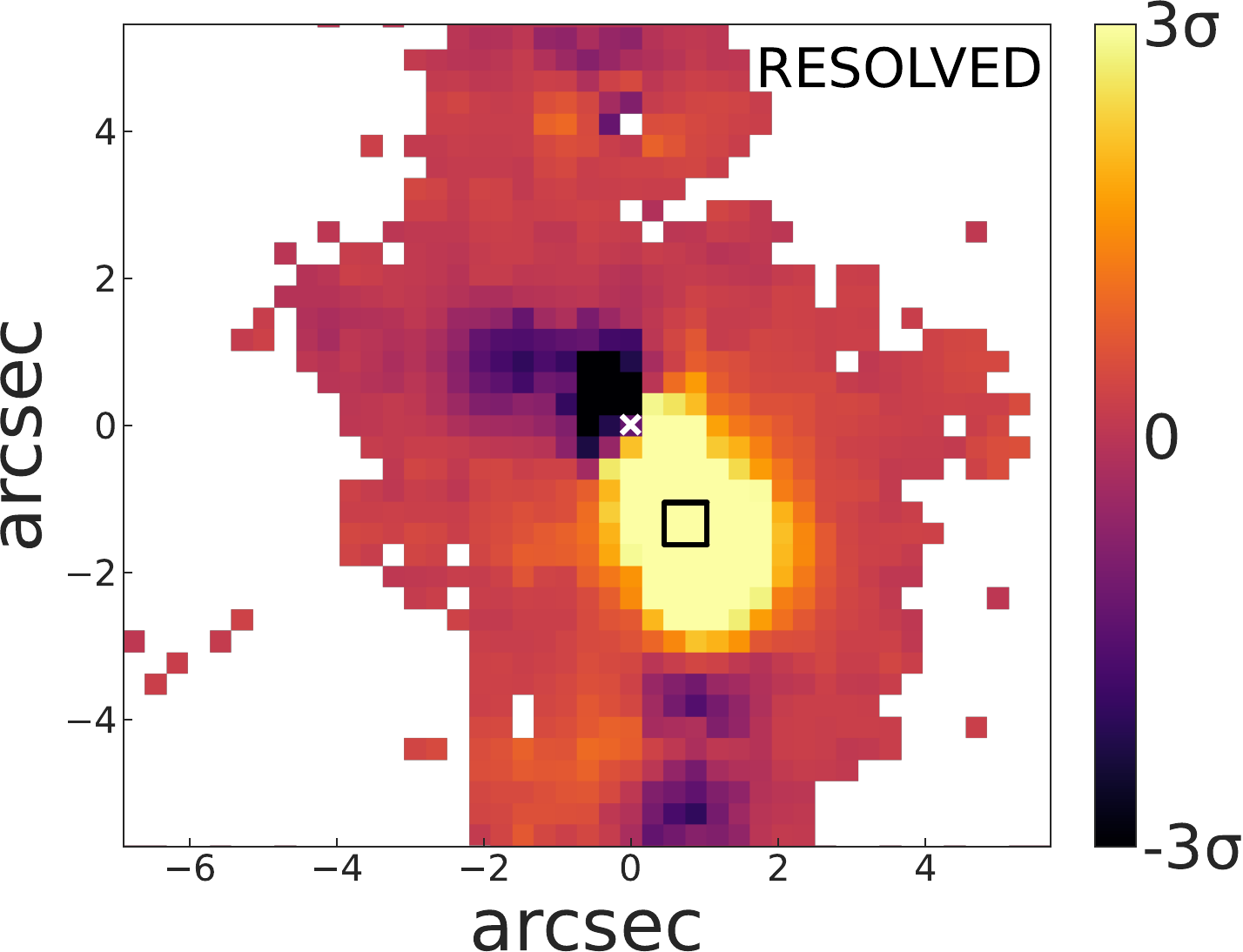} }}
    \qquad
    \subfloat{{\includegraphics[width=4.12cm]{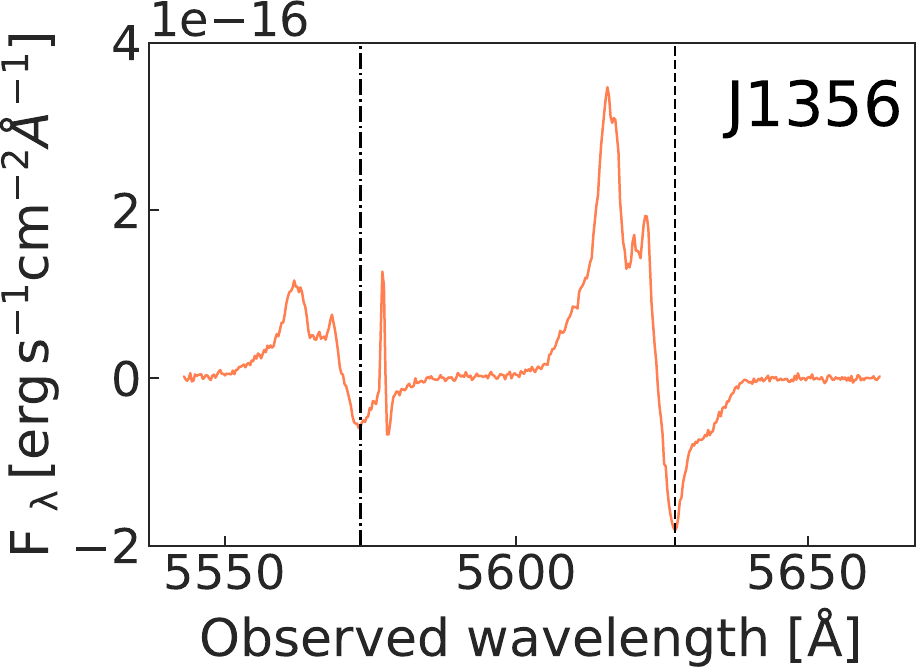} }}
    \qquad
    \subfloat{{\includegraphics[width=4.cm]{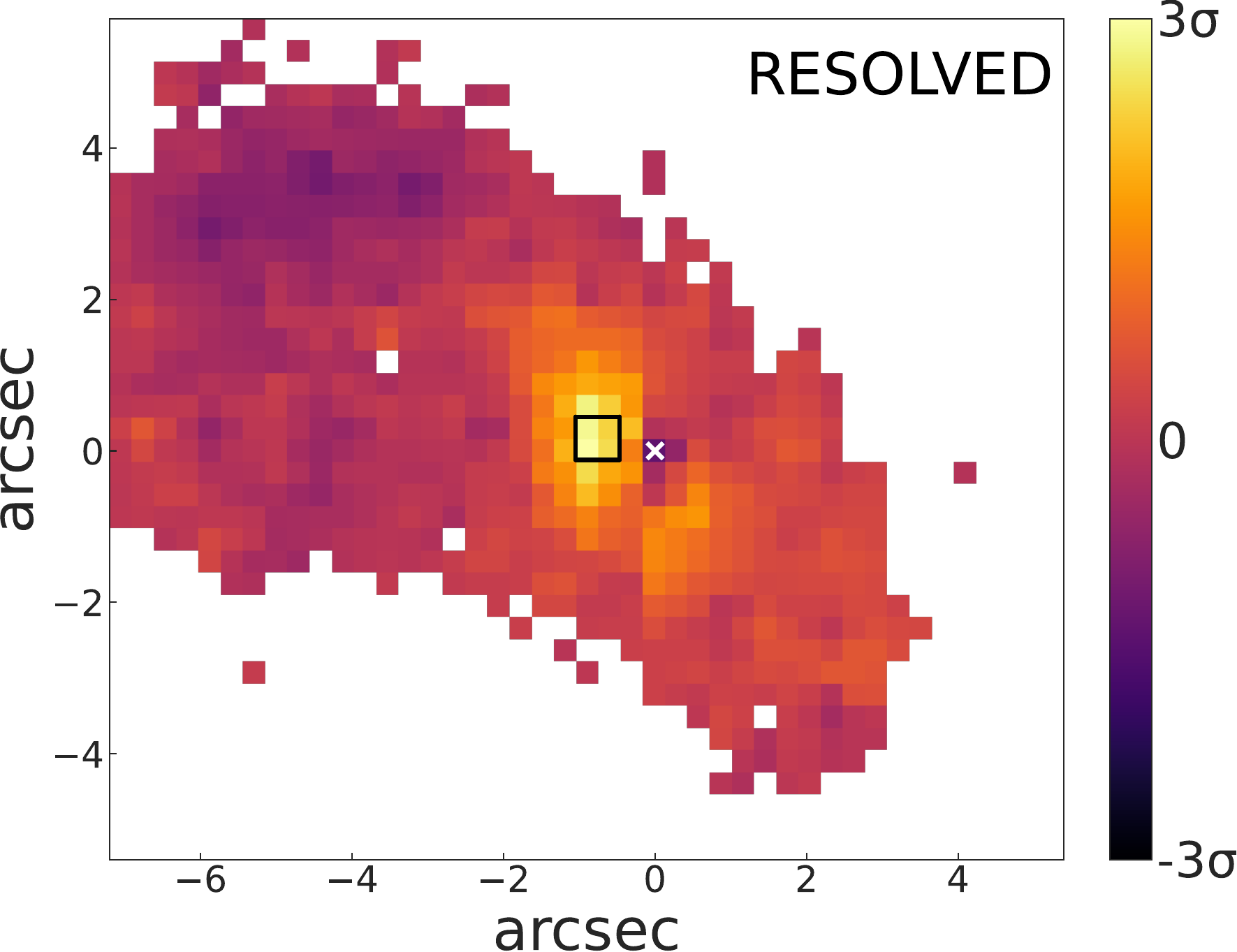} }}
    \qquad
    \subfloat{{\includegraphics[width=4.12cm]{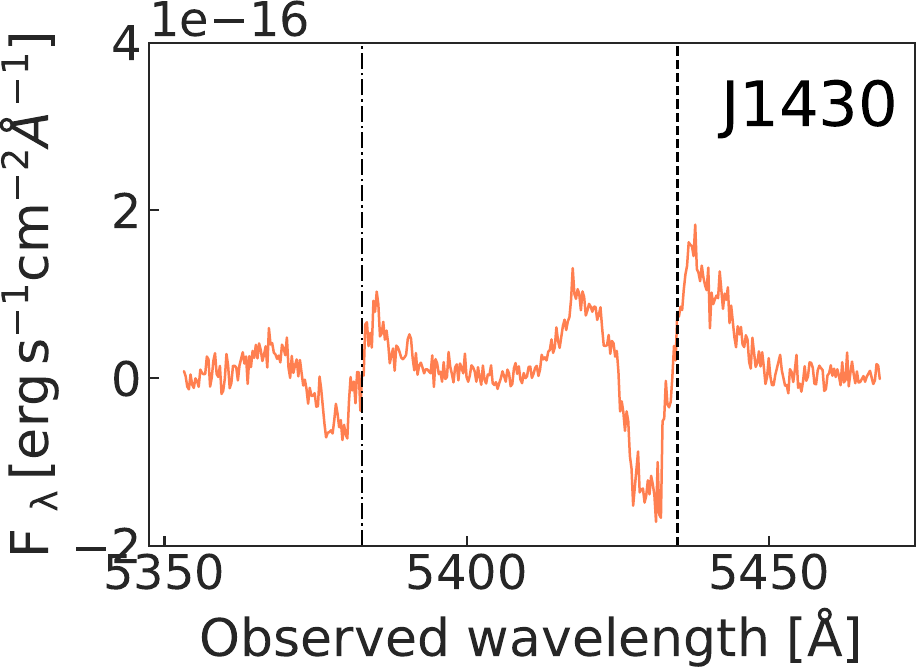} }}
    \qquad
    \subfloat{{\includegraphics[width=4.cm]{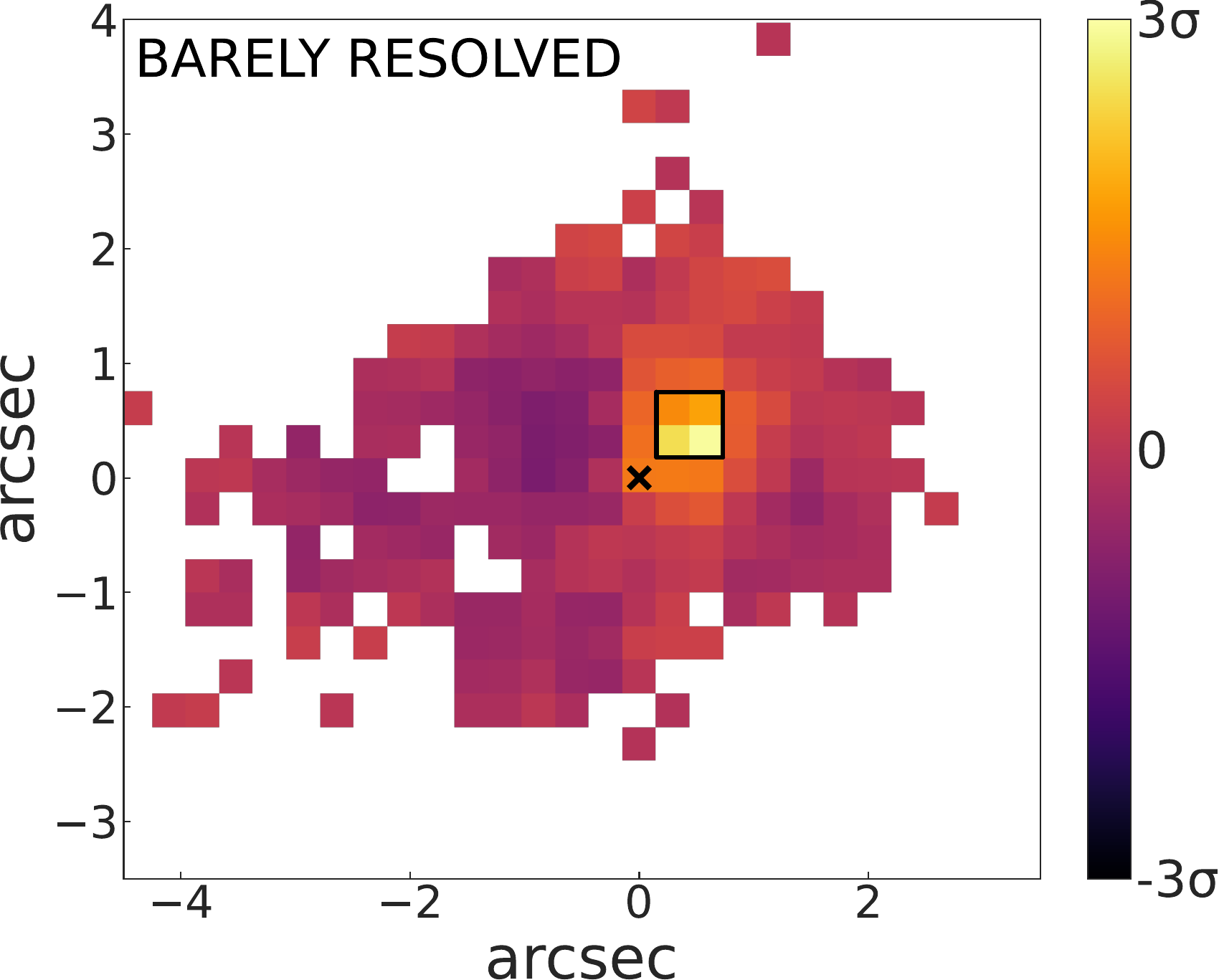} }}
    \qquad
    \subfloat{{\includegraphics[width=4.12cm]{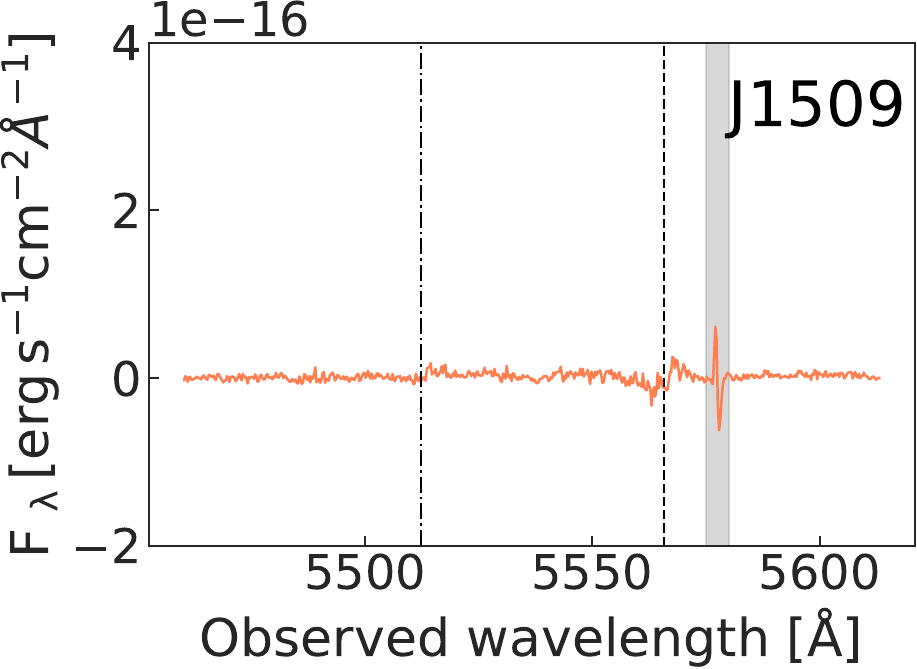} }}

\caption {Residuals obtained from the PSF-subtraction analysis performed across the MEGARA FOV. The left panels show the residual maps generated by subtracting the normalized model of the central spectrum, extracted in the five  spaxels around the peak of the [O~III] emission (black or white crosses), from the spectra of all the spaxels. North is up and east to the left. The black squares indicate the excess regions ($\geq3\sigma$ and in 0.6\arcsec$\times$0.6\arcsec~apertures) from which we extract the corresponding  spectra shown in the right panels. The dot-dashed and the dashed lines correspond to the [O~III]$\lambda$4959 and 5007 \AA~centroids of the central model. For J1509, the grey band in the bottom right panel indicates bad pixels that were excluded when collapsing the spectral range shown on each panel to build the residual maps.
}
\label{fig:PSF}
\end{figure}

 From the analysis of the nuclear spectra presented in Section \ref{nucleus}, we find signatures of ionized outflows in the five QSO2s. However, before exploring the kinematics of the gas across the MEGARA FOV, it is important to evaluate whether the [O~III] emission is spatially resolved or not. In seeing-limited observations such as ours, beam smearing can replicate the nuclear spectrum throughout all the spaxels included within the seeing disc. 
 
 
 To do so we  used the PSF-subtraction method proposed in \citet{Carniani13} and applied in several studies (e.g., \citealt{Carniani15, Kakkad20, Kakkad23}). This  method consists of comparing the observed spectra across the MEGARA FOV with the spectrum extracted at the AGN position. The emission line profiles are expected to change with respect to the central one if the [O~III] emission is spatially resolved. Therefore, we model the spectrum of the central 5  spaxels and we subtract it spaxel by spaxel, after applying a normalization factor to scale the amplitude and to minimize the residuals between the nuclear fit and the observed spectra of the other spaxels. The normalization factor is measured in the entire wavelength range covered by the model (see the right panels of Fig. \ref{fig:PSF}). 
 Before applying this procedure, we masked spaxels having less than three points in the spectrum detected at <3$\sigma$ over the continuum leftwards and redwards of the [O~III] peak. 
Finally, we generate the cube of the residuals collapsing the entire wavelength range shown in the right panels of Fig.~\ref{fig:PSF}  and we inspected them to see whether they are close to zero or show a significant excess. 

 In Fig.~\ref{fig:PSF} we show the residual maps (left panels) and the corresponding spectra (right panels) extracted from the excess regions indicated with black boxes in the corresponding maps. The residual maps are in $\pm3\sigma$ units,  where the sigma corresponds to the  standard deviation of the background measured in the edges of each source's FOV, using an aperture of 1.5$\arcsec \times 1.5\arcsec$. We consider the [O~III] emission resolved when residual structures are observed at $\ge$3$\sigma$. Instead, noisy maps with corresponding low-signal spectra are indicative of unresolved emission. This is the case of J1010,  whose [O~III] emission appears spatially unresolved in our data. Instead, we find it to be spatially resolved in J1100, J1356, J1430, and tentatively resolved in J1509.   Therefore, in the following, we will only consider outflow measurements for the QSO2s where the [O~III] emission is resolved, although in Section \ref{extended} we describe the kinematics maps of J1010 as well.

\subsection{The [O~III] kinematics of the extended emission}
\label{extended} 

\begin{figure*}
\centering
\includegraphics[width=0.49\textwidth]{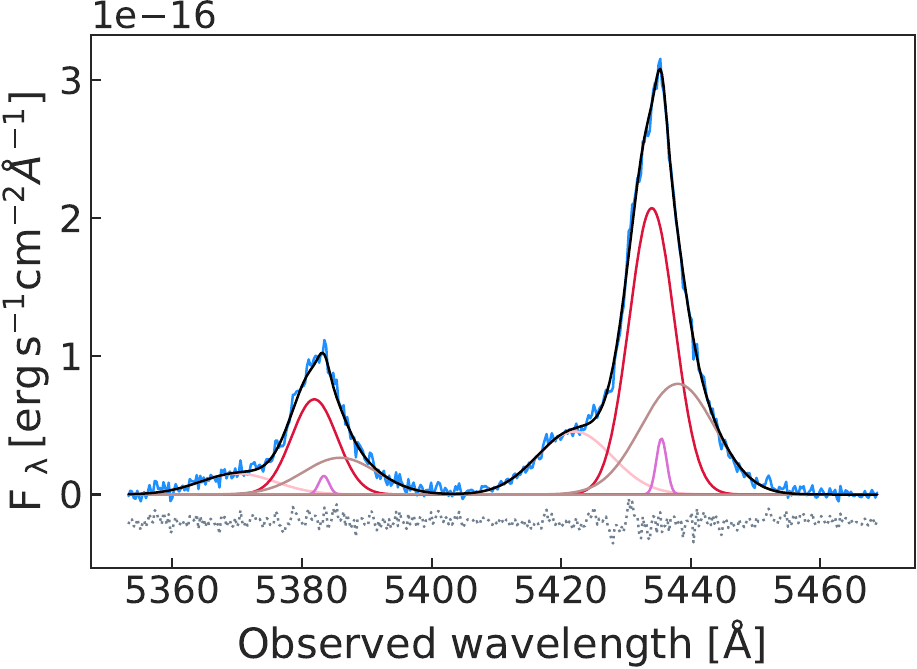}
\includegraphics[width=0.49\textwidth]{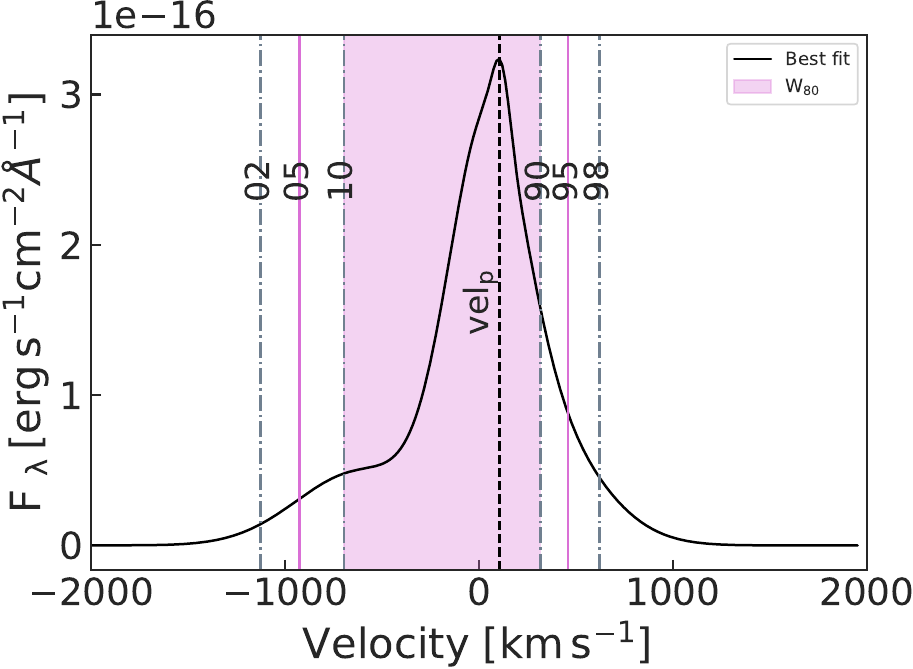}
\caption {Example of non-parametric analysis of a random spaxel of J1430. Left panel: [O~III]$\lambda\lambda$5459,5007 \AA~continuum subtracted emission lines, with the observed spectrum shown in blue, the best fitted model in black, and the corresponding residuals in grey. The same number of Gaussians (four in this case) with the same kinematics and a flux ratio of 2.98 are fitted for each line of the doublet. Right panel: corresponding model of the [O~III]$\lambda$5007 \AA~line, obtained from the fit shown in the left panel, with W$_{80}$ indicated as the pink shaded region, vel$_{05}$ and vel$_{95}$ indicated by the pink vertical lines, and the upper (vel$_{02}$ and vel$_{98}$) and lower bounds (vel$_{10}$ and vel$_{90}$) marked by the dot-dashed vertical gray lines. The dashed vertical black line represents the velocity at the [O~III] peak (vel$_\text{p}$).}
\label{fig:nopar}
\end{figure*}


In order  to characterize the [OIII] kinematics of the QSO2s, we performed a non-parametric analysis of the line profiles by measuring the velocity percentiles, i.e., velocities associated with a percentage of the flux of the emission line (e.g., \citealt{Whittle85, Harrison14, Speranza21, Bessiere22}). We adopt this approach to automatize the fitting procedure across the MEGARA FOV and to overcome the complexity of the line profiles. In fact, some of the QSO2s are part of merging/interacting systems (e.g., J1356) showing a wide diversity of [O~III] profiles at different distances from the AGN, that makes it challenging to obtain an acceptable physical description using a multi-component model. To accomplish the non-parametric approach we fit multiple Gaussians to describe the emission line profiles, as we did in Section~\ref{nucleus}, but without ascribing any physical meaning to the single components. To perform the fits we use the non-linear least-squares minimization and curve fitting for Python (LMFIT, \citealt{Newville14, Newville16}). We fit from a minimum of one to a maximum of six Gaussian components to each line of the [O~III] doublet, including a new Gaussian to the fit only when the reduced $\chi^2$ improves more than 10$\%$ with respect to the previous fit. As for the nuclear analysis presented in Section~\ref{nucleus}, we fix the kinematics and flux ratios (i.e., [O~III]$\lambda$5007/[O~III]$\lambda$4959=2.98) between the lines of the doublet. We set the width of the Gaussians to be larger than the instrumental broadening and to have amplitudes $\geq$ 3$\sigma$, in order to avoid   bad fits.
Once we obtain the best fit for the [O~III] doublet  in a given spaxel, we remove the model of the [O~III]$\lambda$4959 line to obtain an isolated model of the [O~III]$\lambda$5007 emission line. An example of the fitting procedure is shown in the left panel of Fig.~\ref{fig:nopar} (for a random spaxel of J1430), with the corresponding [O~III] model shown in the right panel. In order to avoid spaxels with poor or non detection of the [O~III] line during the fitting process, we excluded those having less than three points in the spectrum detected at <3$\sigma$ over the continuum leftwards and redwards of the emission line peak. This is the same criterion used for the PSF-subtraction analysis described in Section~\ref{PSF}.  

Applying this process spaxel-by-spaxel, we can extract  the parameters used to build the kinematics maps of [O~III]$\lambda$5007. These parameters are shown in the right panel of Fig.~\ref{fig:nopar} and they are the following.

\begin{figure*}
\centering
\includegraphics[width=1.\textwidth]{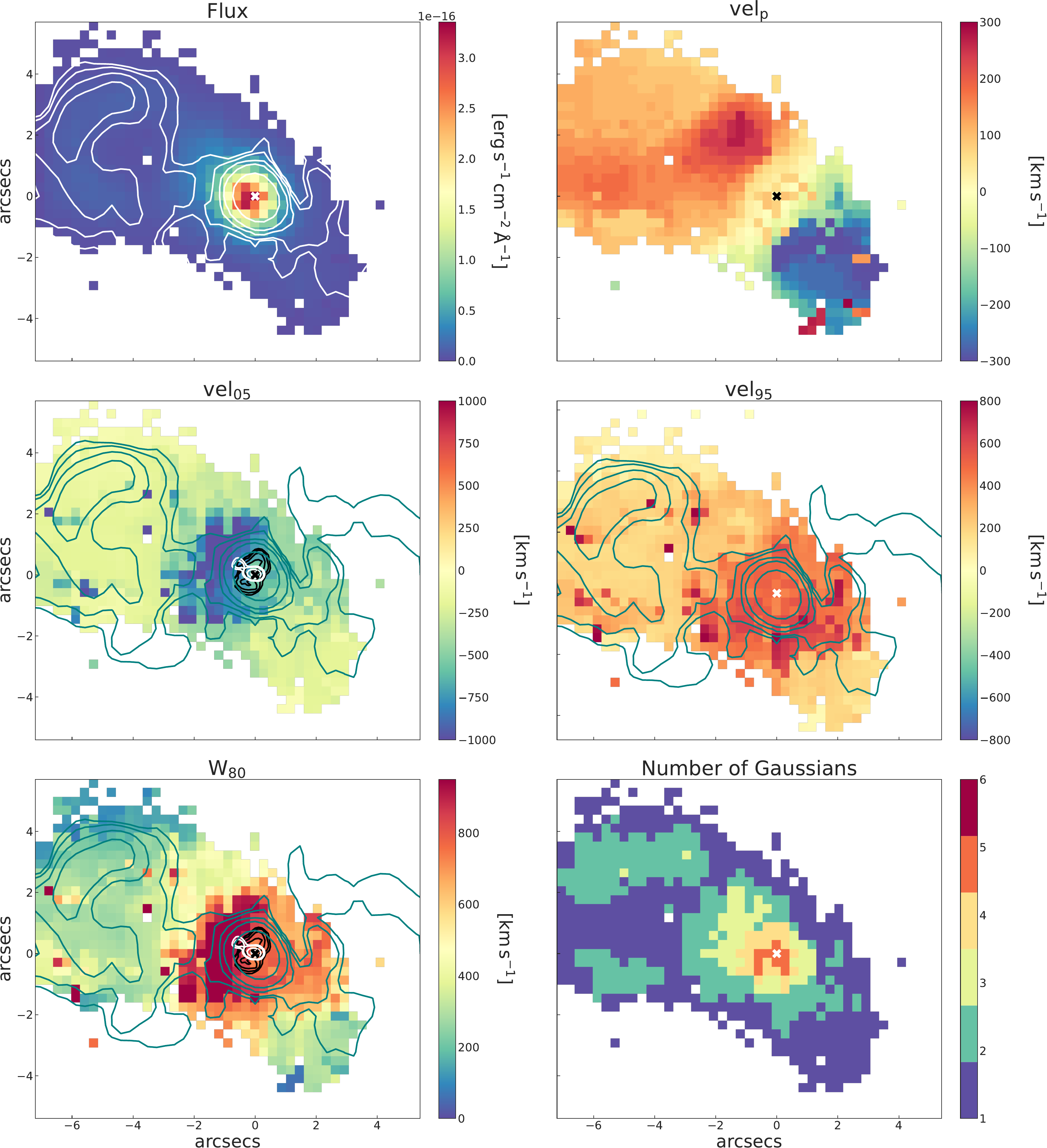}
\captionsetup{width=1.\linewidth}
\caption {Kinematic maps obtained from the non-parametric analysis of the [O~III] emission line profiles of J1430. Top panels: integrated flux on the left, with the $\sim$1\arcsec~resolution 6 GHz VLA contours at 3,5,10,20, and 60$\sigma$ from \citet{Jarvis19} superimposed (white solid lines), and vel$_\text{p}$ on the right, which traces the narrow component. Middle panels: vel$_{05}$ (left) and vel$_{95}$ (right), which trace blueshifted and redshifted outflowing gas. The same 6 GHz contours shown in the flux map are shown in cyan, and the $\sim$0.25\arcsec~resolution 6 GHz VLA contours at 3,5,15,30, and 60$\sigma$ from \citet{Jarvis19} are shown in solid white. The black contours correspond to the brightness temperature ratio (T$_{32}$/T$_{21}$) from ALMA CO observations at 0.5\arcsec ~resolution (\citealt{Audibert23}). Bottom panels: W$_{80}$ on the left, with the same contours as in the vel$_{05}$ map, and the number of Gaussians fitted in each spaxel on the right. White and black crosses indicate the AGN location, defined as the maximum of the [O~III] emission. North is up and east to the left.}
\label{fig:J1430_map}
\end{figure*}

\begin{itemize}
   \item The velocity peak (vel$_p$), which is the velocity shift of the peak of the line from the systemic velocity measured in the nuclear spectrum (i.e., from $\lambda_c$ in Table~\ref{tab:nuc}). vel$_p$   mainly
    traces the narrow component, as can be seen from the top right panel of Fig. \ref{fig:J1430_map}. 
   \item The 5th and 95th velocity percentiles (vel$_{05}$ and vel$_{95}$), which are the velocities including the 5$\%$ and 95$\%$ of the emission line flux. They are representative of the blueshifted and redshifted outflow velocities, respectively (see middle panels of Fig. \ref{fig:J1430_map}).
   \item W$_{80}$, which is the line region including 80\% of the flux (W$_{80}$ = vel$_{90} - $vel$_{10}$). This parameter is indicative of how turbulent the gas is and, in the case of a line profile that can be described with a single Gaussian, it approximately corresponds to  $1.09 \times$ FWHM (see bottom left panel of Fig. \ref{fig:J1430_map}).   We note that the values of W$_{80}$ have not been corrected from instrumental broadening.
\end{itemize}

In Fig.~\ref{fig:J1430_map} we present the kinematic maps of J1430, and the rest of the maps are presented and described in Appendix~\ref{AppendixA} (see Figs.~\ref{fig:J1010_extended}-\ref{fig:J1509_extended}). The top left panel of Fig.~\ref{fig:J1430_map} shows the [O~III]$\lambda$5007 integrated flux. Due to the peculiar morphology of the ionized gas, J1430 has been nicknamed the Teacup (\citealt{keel12}).  This  morphology has been also detected  in the radio (\citealt{Harrison15}) and  in the X-rays (\citealt{Lansbury18}). The ionized bubble reaching 12 kpc north-east from the nucleus is detected in our map, as well as part of the south-west structure. As shown in the top right panel,  negative velocities are seen to the south-west and positive velocities to the north-east, up to $\pm$300\kms. This is in agreement with previous results (\citealt{Gagne14,Harrison14,Ramos17,Villar18,Moiseev23, Venturi23}). In the case of this QSO2, strong jet-ISM interaction is inducing enhanced turbulence in both the molecular and ionized gas, and the ionized and molecular major axes do not coincide (\citealt{Audibert23,Venturi23}), possibly due to the past merger even that J1430 experienced. Recently, \citealt{Moiseev23} performed kinematic modelling of the ionized gas and claimed that most of it is rotating. However, another possibility is that the gas that we are tracing with vel$_p$ is part of the outflow itself (see below).


The middle panels of Fig.~\ref{fig:J1430_map} show high values of  blueshifted and redshifted outflow velocities in the north-east and south-west direction respectively, with vel$_{05}$ of up to $\sim$-1000 \kms and vel$_{95}$ of up to $\sim$800 \kms. These high velocities are detected within the same region where W$_{80}$ increases over $\sim$600\kms~(see the left bottom panel of Fig.~\ref{fig:J1430_map}). The maximum values of W$_{80}$ are co-spatial with the most blueshifted velocities (i.e., to the north-east). This is in agreement with \citet{Harrison14, Harrison15} and \citet{Ramos17}.
\citet{Keel17} reported blue [O~III] wings eastwards of the nucleus and red wings westwards, with velocities of up to $\pm$1000 \kms, using GMOS IFS observations. Thanks to the higher spectral resolution of our data, to the non-parametric analysis we performed, and to the larger FOV of MEGARA, which is more than twice as big as the FOV of the GMOS IFU data used by \citet{Harrison14} and \citet{Keel17}, we clearly resolve the approaching and receding sides of the outflow in J1430, in the form of expanding shells/bubble. It is also possible that the reverse velocity pattern that we see in the vel$_p$ map of J1430 corresponds to the inner and outer sides of north-east and south-west hollow cones, respectively, but here we do not perform any kinematic modelling that might allow us to confirm this.  The last panel of Fig.~\ref{fig:J1430_map} shows the number of Gaussians fitted in each spaxel,  which varies from five with increasing distance from the nucleus.

From the other maps presented in Appendix~\ref{AppendixA},  which are described there, we detect  gas rotation in  in J1010, J1100, and J1509, based on the similarity with the CO gas distribution (\citealt{Ramos22}).   In J1356 and J1430, the kinematics are clearly affected by the mergers, which increase the turbulence. J1100 shows disturbed rotation, which might be related to the presence of the stellar bar (\citealt{Fischer18,Ramos22}). J1430 and J1356 are the only objects in the sample in which we observe the approaching and receding sides of the outflows, with both QSO2s having similar outflow velocities of $\pm$800 \kms. However, in J1356, the distribution of the outflowing gas is irregular, showing the highest blueshifted and redshifted velocities and W$_{80} \approx$  1000 \kms~in the south-west region (PA$\sim 200^{\circ}$). For the rest of the targets, we only detect negative outflow velocities (vel$_{05}$) coinciding with high velocity dispersion (W$_{80}$) regions. In J1100 the most turbulent gas is observed in the north-east (PA$\approx 60^{\circ}$), where W$_{80}\sim$ 1600-1800 \kms~and vel$_{05} \geq -1500$ \kms. In J1509, the highest values of W$_{80}$, of 1600-1700 \kms~and vel$_{05}$, of up to -1500 \kms, are measured in the north-west. Finally, for J1010 we detect  high values of vel$_{05}$ and vel$_{95}$, but similar values of W$_{80}$ are measured across almost the entire extent of the [O~III] emission-line region, consistent with this being unresolved (see Section~\ref{PSF}). Despite this, we detect a rotation pattern in the vel$_p$ map of J1010 because approximately half of the instrument fibers are probing redshifted velocities and the other half blueshifted velocities. In cases of severe beam smearing, the amplitude of rotation is usually smaller than the real one. 

\subsection{Outflow properties}
\label{properties}

\begin{table*}
\caption{Outflow properties measured from the [O~III]$\lambda 5007 \AA$ emission line.}
\centering
\begin{tabular}{c c c c c c c c c}
\hline
ID   &   log n$_{\rm e}$*      &  M$_{\text{out}} \times 10^{6}$ &  R$_{\text{out}}$ & PA & v$_{\text{out}}$ &  $\dot{\text{{M}}}_{\text{{out}}}$  &   log $\dot{\text{{E}}}_{\text{{kin}}}$  & $\xi$  \\
            & $[\text{cm}^{-3}]$       &   $[\text{M}_{\odot}]$  & [kpc] & $[^{\circ}]$ & $[\text{km}~\text{s}^{-1}]$  &   $[\text{M}_{\odot}~\text{yr}^{-1}]$ &   $[\text{erg}~\text{s}^{-1}]$ &  [\%]  \\
            
     (1)   &  (2)  &   (3) &  (4)  &  (5) &    (6) &   (7)&   (8)  & (9) \\
\hline
J1100  & $(3.0^{+0.09}_{-0.09})_{\text{[S~II]}}$ &   $8.72^{+2.00}_{-1.63}$  & $5.1^{+0.2}_{-0.3}$ & 63 & $-1236^{+358}_{-441}$ & $6.5^{+1.5}_{-1.2}$ & 42.02 & 0.0148 \\
J1356  & $(2.51^{+0.1}_{-0.1})_{\text{[S~II]}}$  &   $35.01^{+6.82}_{-5.40}$  & $12.6^{+0.1}_{-0.5}\,\vert\,6.8^{+3.7}_{-1.8}$ &  197$\,\vert\,$200 & -631$^{+145}_{-190}\,\vert\,483^{+178}_{-135}$ & $6.1^{+1.1}_{-0.9}$ &  41.32 & 0.0060 \\
J1430 & $(3.01^{+0.18}_{-0.18})_{\text{[S~II]}}$   &    $4.64^{+1.73}_{-1.14}$   & $3.7^{+0.4}_{-0.2}\,\vert\,3.1^{+0.3}_{-0.3}$ & 65$\,\vert\,$198 & -760$^{+226}_{-254}\,\vert\,$529$^{+217}_{-159}$  & $3.3^{+1.0}_{-0.7}$ &  41.13 & 0.0020 \\
J1509 & $(2.85^{+0.1}_{-0.1})_{\text{[S~II]}}$   &    $4.00^{+1.04}_{-0.83}$   & $3.8^{+0.3}_{-0.2}$   & 320 & -1289$^{+274}_{-317}$ & $4.2^{+1.0}_{-0.8}$ & 41.18 & 0.0012 \\
\hline

ID   &   log n$_{\rm e}$**     &  M$_{\text{out}} \times 10^{6}$ &  R$_{\text{out}}$ & PA & v$_{\text{out}}$ &  $\dot{\text{{M}}}_{\text{{out}}}$  &   log $\dot{\text{{E}}}_{\text{{kin}}}$  & $\xi$  \\
            & $[\text{cm}^{-3}]$       &   $[\text{M}_{\odot}]$  & [kpc] & $[^{\circ}]$ & $[\text{km}~\text{s}^{-1}]$  &   $[\text{M}_{\odot}~\text{yr}^{-1}]$ &   $[\text{erg}~\text{s}^{-1}]$ &  [\%]  \\
\hline
J1100  & $(3.99^{+0.07}_{-0.08})_{\text{TA}}$ &   $0.89^{+0.16}_{-0.15}$  & $5.1^{+0.2}_{-0.3}$ & 63 & $-1236^{+358}_{-441}$  & $0.7^{+0.1}_{-0.1}$ & 41.03 & 0.0015 \\
J1356  & $(3.21^{+0.00}_{-0.15})_{\text{TA}}$  &    $6.99^{+0.23}_{-1.52}$  & $12.6^{+0.1}_{-0.5}\,\vert\,6.8^{+3.7}_{-1.8}$ &  197$\,\vert\,$200 & -631$^{+145}_{-190}\,\vert\,483^{+178}_{-135}$ &  $1.2^{+0.1}_{-0.2}$ &  40.58 & 0.0011 \\
J1430  & $(3.24^{+0.05}_{-0.3})_{\text{TA}}$  &    $2.73^{+0.25}_{-0.99}$  & $3.7^{+0.4}_{-0.2}\,\vert\,3.1^{+0.3}_{-0.3}$ & 65$\,\vert\,$198 & -760$^{+226}_{-254}\,\vert\,$529$^{+217}_{-159}$  & $1.6^{+0.1}_{-0.6}$ & 40.90 & 0.0011 \\
J1509 & $(3.41^{+0.11}_{-0.21})_{\text{TA}}$  & $1.10^{+0.32}_{-0.42}$    & $3.8^{+0.3}_{-0.2}$   & 320 & -1289$^{+274}_{-317}$ & $1.1^{+0.3}_{-0.4}$ & 40.61 & 0.0003 \\
\hline
\end{tabular}
\tablefoot{(1) Object ID; (2) electron density; 
(3) mass of the outflowing gas;
(4) maximum outflow radius;
(5) position angle of the outflow;
(6) outflow velocity measured from the velocity maps (negative velocities correspond to vel$_{05}$ and positive to vel$_{95}$); 
 (7) mass outflow rate computed as 3$\,\times\,\text{v}_{\text{out}}\times \text{M}_{\text{out}}$/R$_{\text{out}}$;  
 (8) outflow kinetic energy calculated as 0.5$\times\dot{\text{M}}_{\text{out}} \times(\text{v}_{\text{out}}^{2} + 3\sigma^2)$, with $\sigma$ being the velocity dispersion, measured as FWHM = 2.355/$\sigma$, with FHWM = W$_{80}/1.09$; 
 (9) coupling efficiency ($\xi$= $\dot{\text{{E}}}_{\text{{kin}}}/\text{L}_{\text{bol}}$). 
 *Electron densities computed from the [S~II]$\lambda\lambda 6716,6731$ doublet  ([S~II]). **Electron densities computed using the trans-auroral lines  (TA). Two values of R$_{\text{out}}$, PA, and v$_{\text{out}}$ are reported for J1356 and J1430 because the redshifted side of the outflow is also detected. For them, the outflow mass rates, kinetic powers, and coupling efficiencies are calculated independently for the blueshifted and redshifted sides of the outflow, and then added.}
\label{tab:out_prop}
\end{table*}

 From the analysis of the [O~III] emission and kinematics presented in Sections \ref{PSF} and \ref{extended}, we concluded that they are spatially resolved in J1100, J1356, J1430, and J1509. Therefore, in the following we describe and measure the outflow properties of these QSO2s, because our data do not allow us to constrain the outflow properties of J1010.

The velocity maps shown in Figs. \ref{fig:J1430_map}, \ref{fig:J1100_extended}, \ref{fig:J1356_extended}, and \ref{fig:J1509_extended} reveal different patterns: J1430 shows an expanding shell/bubble with negative and positive velocities, J1100 and J1509 present high blueshifted velocities coincident with the highest W80 values, and J1356 shows an irregular and complex gas morphology having positive and negative outflow velocities. Therefore, we need to analyse each target separately to estimate the outflow properties and their impact on the surrounding environment, as well as the mechanism(s) that might be driving them. 


We first need to measure the electron densities (n$_e$) and outflow fluxes to estimate the outflow masses (M$_{\text{out}}$), and the outflow radii (R$_{\text{out}}$) and velocities (v$_{\text{out}}$) to quantify the mass outflow rates ($\dot{\text{{M}}}_{\text{{out}}}$) and kinetic powers ($\dot{\text{{E}}}_{\text{{kin}}}$) of the winds. 
The largest uncertainty is given by n$_e$, which can change the results by several orders of magnitude (e.g., \citealt{Rose18,Davies20,2023MNRAS.tmp.1680H}). To evaluate how $\dot{\text{{M}}}_{\text{{out}}}$ varies with n$_e$, we measure the outflow density using two different methods involving different emission lines detected in the SDSS spectra available for the QSO2s. 
To measure the line fluxes we fitted a first-order polynomial to the continuum and a single Gaussian profile to each line.  A multiple Gaussian fit including broad and/or intermediate components was not done because of the low signal-to-noise of some of the emission lines involved. Therefore, the two sets of densities correspond to total densities,  and they are most likely lower limits to the outflow density (e.g., \citealt{Mingozzi19, Fluetsch21}).

\begin{figure*}
\centering
\includegraphics[width=0.33\textwidth]{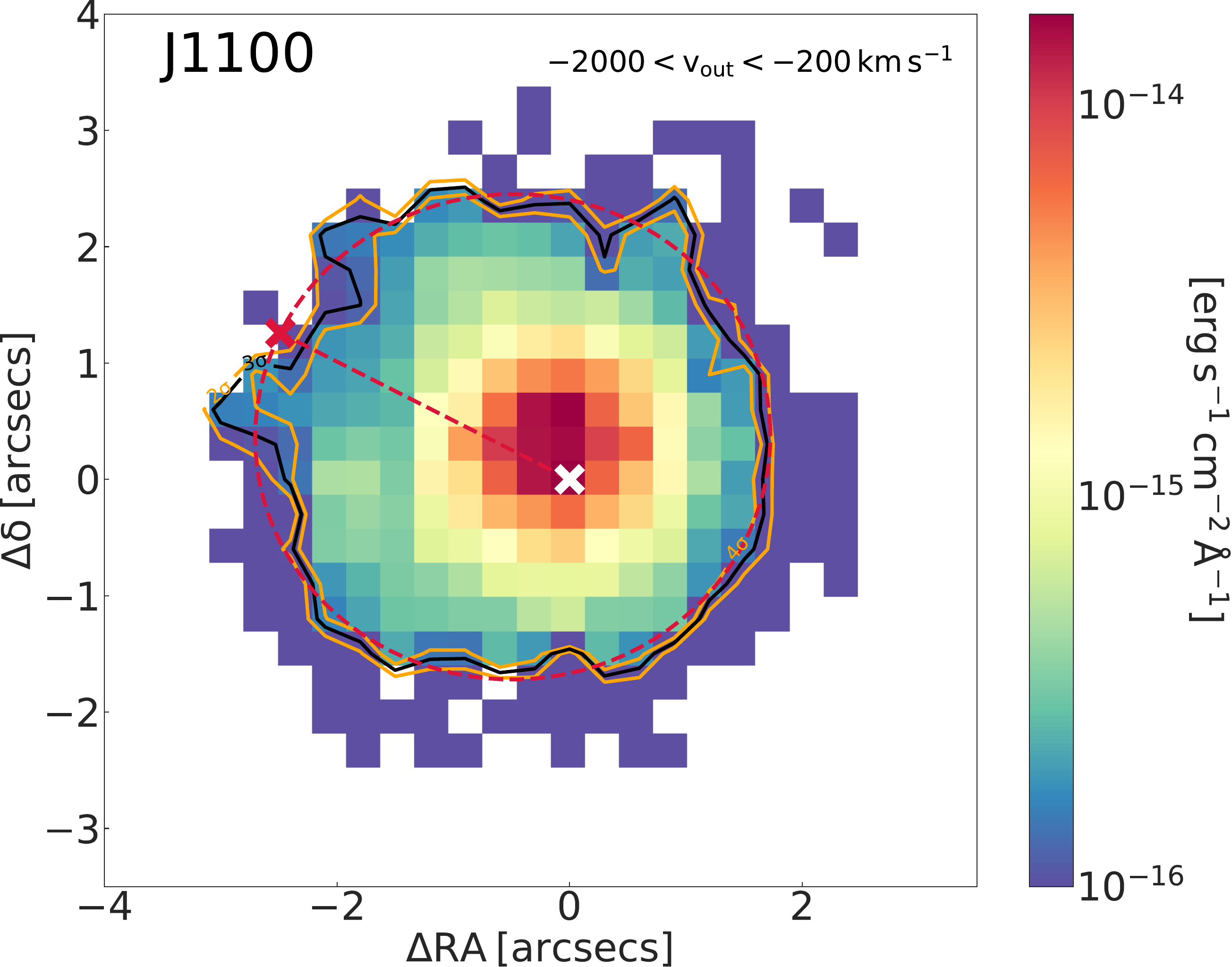}
\includegraphics[width=0.33\textwidth]{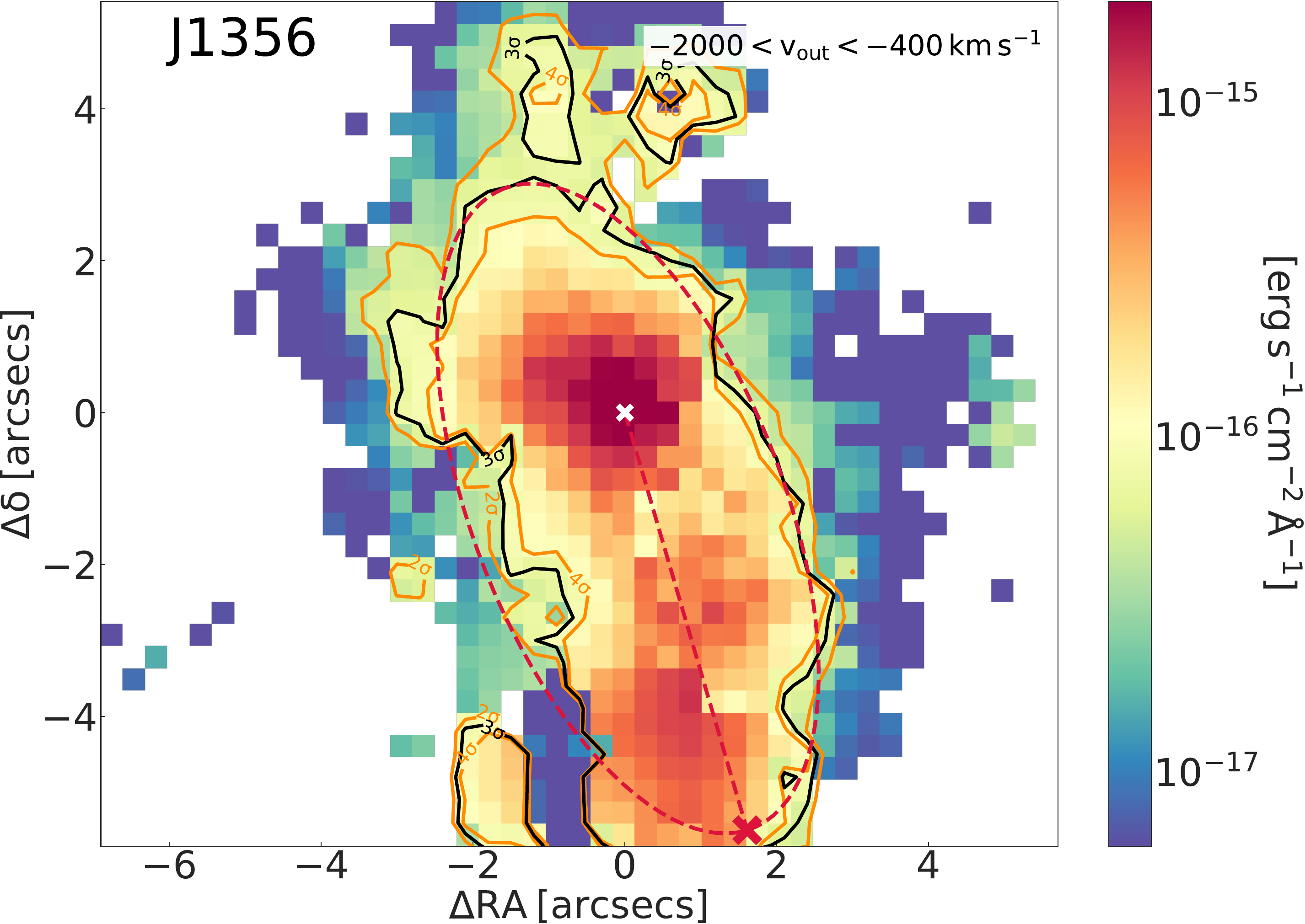}
\includegraphics[width=0.33\textwidth]{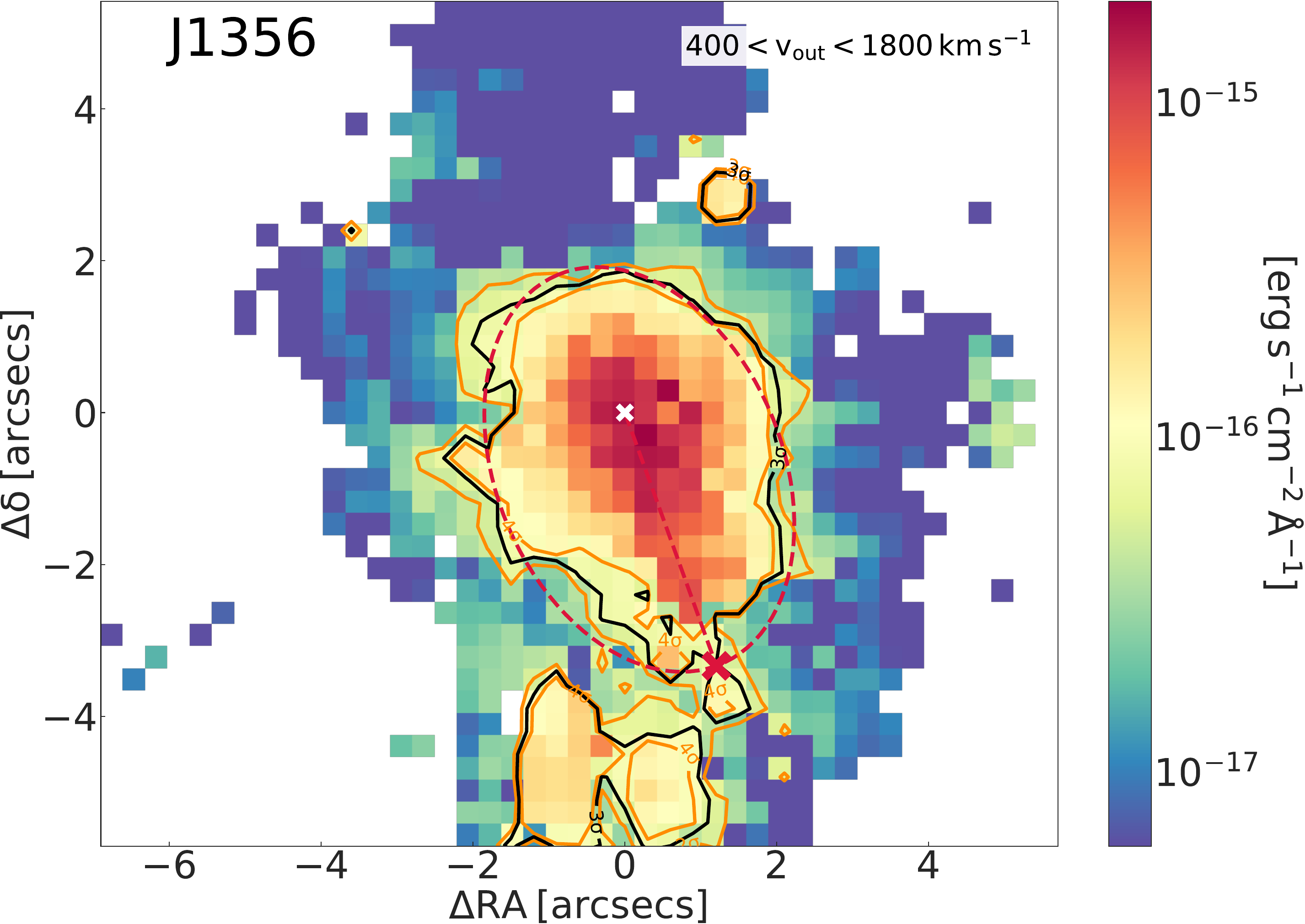}
\includegraphics[width=0.33\textwidth]{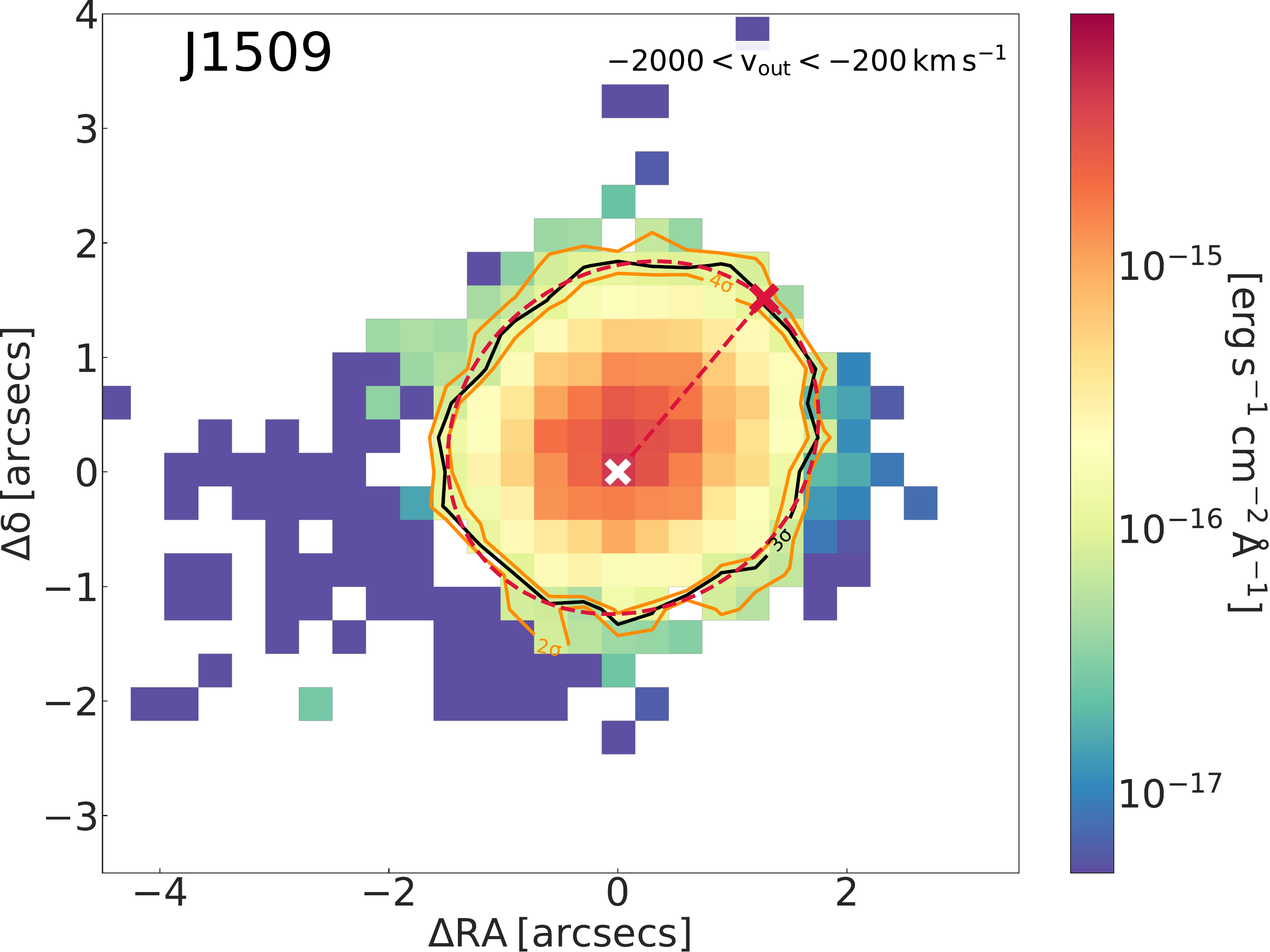}
\includegraphics[width=0.33\textwidth]{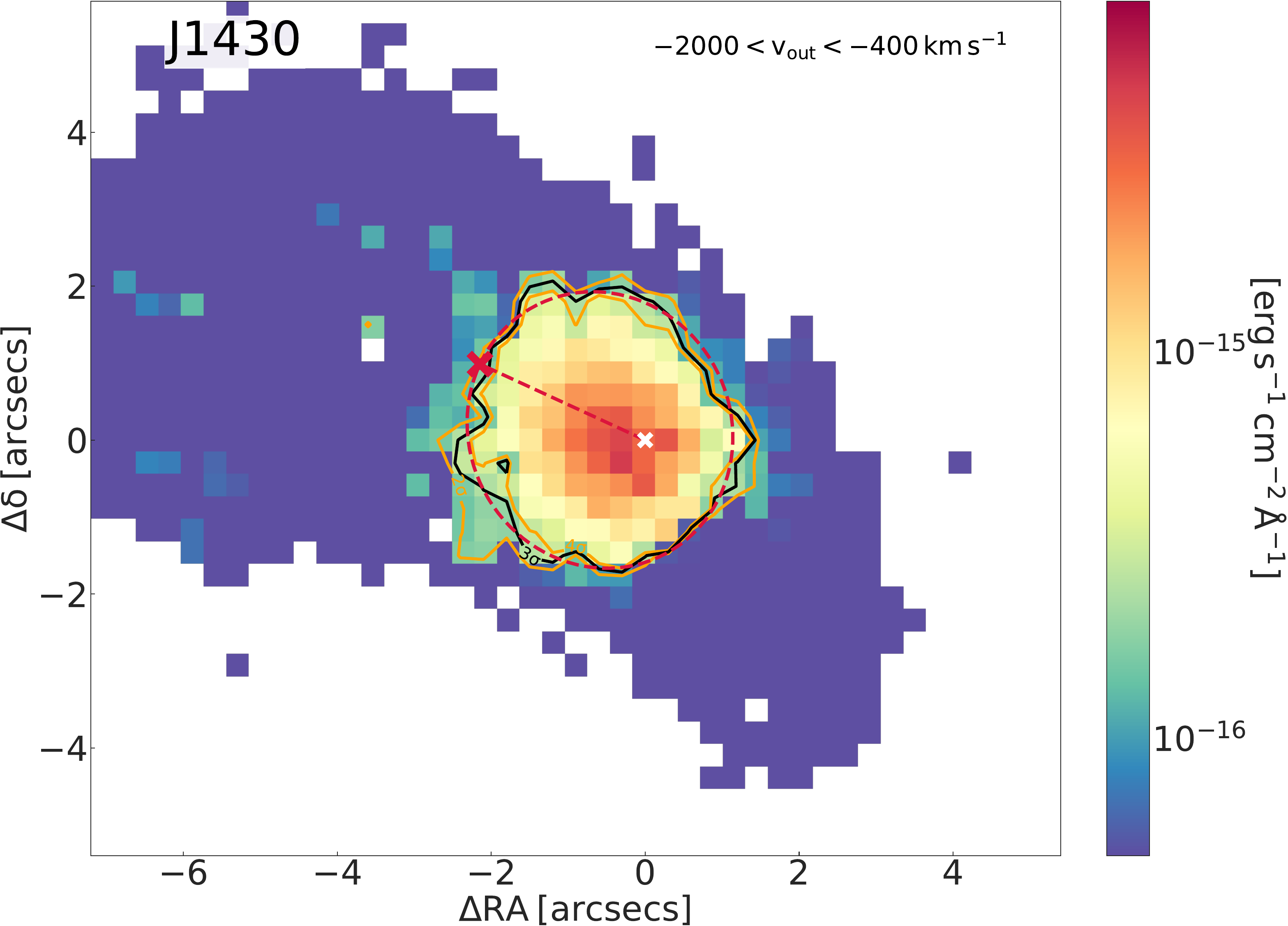}
\includegraphics[width=0.33\textwidth]{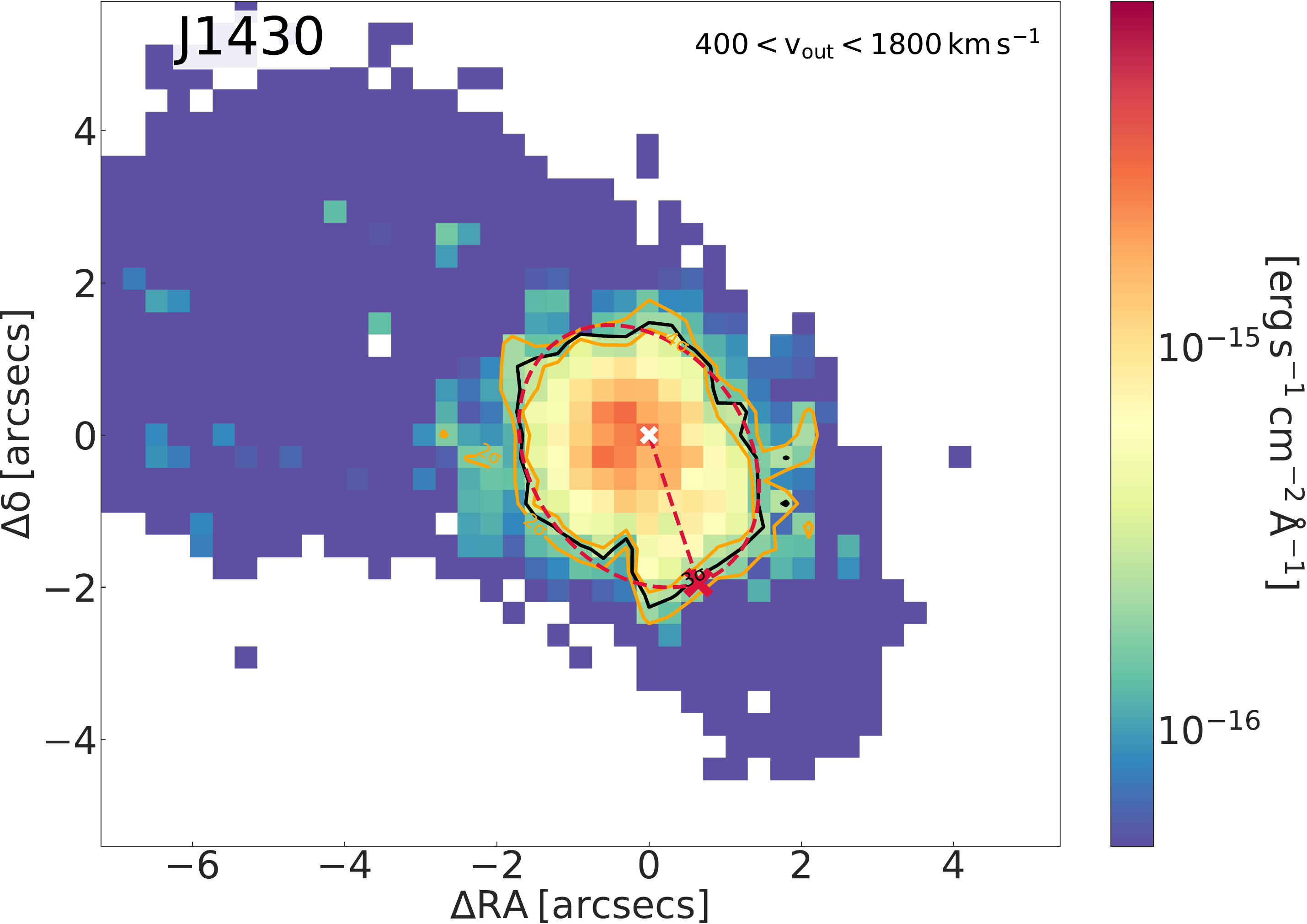}

\caption {[O~III] flux maps corresponding to high-velocity gas only, with velocities between -2000 < v$_{\text{out}}$ < - 200\kms, -2000 < v$_{\text{out}}$ < - 400\kms, and 400 < v$_{\text{out}}$ < 1800 \kms~depending on the source. 
The black solid line corresponds to the flux contour at 3$\sigma$ and the red dashed line to the ellipse fitted to it. The white cross indicates the [O~III] emission peak, and the red cross corresponds to the most distant point from the white cross within the fitted ellipse. The distance between the two crosses is the maximum outflow radius (R$_{\text{out}}$) reported in Table \ref{tab:out_prop}. The two orange lines are the flux contours at the 2$\sigma$ and 4$\sigma$ levels used to estimate the errors of the outflow radius and flux. }
\label{fig:out_r}
\end{figure*}

The first method is based on the density-sensitive ratio of [S~II]$\lambda\lambda$6716,6731 \AA~(\citealt{Osterbrock06}) and the temperature-sensitive ratio of [O~III]$\lambda\lambda$4363,5007 \AA. Using these lines  and the {\it Pyneb} tool (\citealt{Luridiana15}) we obtain densities between 2.5 $\lesssim$ log n$_{\rm e} \lesssim$ 3 cm$^{-3}$ (see Table~\ref{tab:out_prop}). However, this classical technique suffers for saturation at high densities (log n$_e\geq3.5\,\text{cm}^{-3}$; \citealt{Rose18}) and it could lead to an overestimation of the outflow masses and, consequently, of $\dot{\text{{M}}}_{\text{{out}}}$. To overcome this problem the trans-auroral lines can be used instead (\citealt{Holt11,Rose18,Ramos19,2023MNRAS.tmp.1680H, Holdenb}). This second method makes use of the [S~II]$\lambda\lambda$6716,6731 \AA~and [O~II]$\lambda\lambda$3726,3729 \AA~doublets as well as of the trans-auroral [O~II]$\lambda\lambda$7319,7331 \AA~and [S~II]$\lambda\lambda$4068,4076 \AA~doublets.  The electron densities are derived by comparing the [O~II] and [S~II] ratios TR([O~II]) = F(3726+3729)/F(7319+7331) and TR([S~II]) = F(4068+4076)/F(6717+6731) with a grid of photoionization models computed with Cloudy (\citealt{Ferland13}), using a spectral index of $\alpha = 1.5$ and a ionisation parameter of log U=-2.3. 
The resulting densities obtained from the trans-auroral lines are 3.2 $\lesssim$ log n$_{\rm e} \lesssim$ 4 cm$^{-3}$. We list the individual values obtained with both methods in Table~\ref{tab:out_prop}, and for further details on the estimations of the electron densities we refer the reader to Section~3.2 of \citet{Speranza22}, where the two methods are deeply explained. 

To measure the outflow flux we integrate the [O~III] outflow emission at $\geqslant$ 3$\sigma$ (black contours in Fig.~\ref{fig:out_r}), as in \citet{Speranza22}. To do so, we selected the high-velocity [O~III] gas, which varies from case to case: $-$200 and $-$2000 \kms ~for J1100 and J1509, and $-$400 and $-$2000\kms ~and 400 and 1800\kms ~for J1356 and J1430, for which the redshifted counterpart of the outflow is also detected. The interval of velocities are relative to vel$_p$ of each spaxel. 
Since here we are using a non-parametric analysis, the flux is computed using the modelled [O~III]$\lambda 5007 \,\AA$ profile (see Fig.~\ref{fig:nopar} for reference). This has the advantage that we do not have to worry about other emission lines close to [O~III]$\lambda 5007 \,\AA$ or noisy continua,  making it possible to integrate the flux at large velocities.  Once we have the outflow flux maps, we fit an ellipse to the 3$\sigma$ contour (red dashed lines in Fig.~\ref{fig:out_r}) and we integrate the flux within the ellipse. The ellipse is fitted using a Python script presented by \citet{Hill16} in the online supplementary material. 
We calculate the errors on the flux using ellipses fitted to the 4$\sigma$ and 2$\sigma$ contours, shown as orange lines in Fig.~\ref{fig:out_r}. The [O~III] fluxes are then corrected from extinction using the A$_v$ values reported by \citet{Kong18} for our QSO2s  and the extinction law of \citet{1989ApJ...345..245C}. Using the previously described outflow densities and fluxes, we can  measure the corresponding outflow masses using Equation~5 in \citet{Carniani15} (see also \citealt{Osterbrock06}):
\begin{equation}
\centering
\text{M}_{\rm out} = 4\times 10^7 \text{M}_{\odot}  \left( \frac{\text C}{10^{\text{O}/\text{H}}}\right) \left(\frac{\text{L}_{\text{out}}}{10^{44}}\right) \left(\frac{<\text{n}_e>}{10^{3}}\right)^{-1}
 \end{equation}
where L$_{\text{out}} = 3 \times \text{L}_{\text{[O~III]}}$ to consider the total mass of ionized outflow as in \citet{Fiore17}, n$_e$ is the electron density, C=<n$_e^2$>$-$<n$_e$>$^2=1$ is the condensation factor, and O/H=[O/H]-[O/H]$_{\odot} = 0$ (assuming that the gas clouds have the same electron density and solar metallicity as in \citealt{Bischetti17}).

To measure the outflow radius (R$_{\text{out}}$) we use the same outflow flux maps described above. We obtain R$_{\text{out}}$ as the maximum distance between the  [O~III] emission peak and the ellipse that we fitted to the 3$\sigma$ contours (see Fig. \ref{fig:out_r}). By doing this we simultaneously obtain the outflow radius and PA, measured from the north. This is the same method proposed in \citet{Speranza22}, which allows us to include spaxels with low surface brightness but high values of velocity and velocity dispersion that would be missed by measuring the FWHM of the high-velocity flux maps (as in \citealt{Ramos17}). Here, we also estimate the errors by measuring the corresponding extensions at 4$\sigma$ and 2$\sigma$ (see Fig.~\ref{fig:out_r}).   
Following \citet{Kang18}, we then subtract in quadrature the half width at half maximum (HWHM) measured from the standard stars observed after each target,  which is half of the seeing FWHM measured for each QSO2 (shown in Table \ref{tab:obs}), from the outflow radius. With this method, we measure a range of  R$_{\text{out}}$ values for the QSO2s, from 2$\arcsec$ ($\sim$ 3.1~kpc) to 5.7$\arcsec$ ($\sim$ 12.6~kpc). For comparison, \citet{Ramos17,Ramos19} reported radial sizes of 1.1$\pm$0.1 kpc and 1.3$\pm$0.2 kpc for the ionized outflows detected  in J1430 and J1509, respectively, whereas here we measure R$_{\text{out}}$=3.7$\pm^{0.4}_{0.2}$ kpc for the blueshifted side of the outflow in J1430 and R$_{\text{out}}$=3.8$\pm^{0.3}_{0.2}$ for J1509.

Finally, to measure the outflow velocity within the outflow radius, we averaged the values of vel$_{05}$ and vel$_{95}$ (if the redshifted outflow counterpart is detected) from the outflow velocity maps (see middle panels of Fig.~\ref{fig:J1430_map}). We did this by selecting only those spaxels within the 3$\sigma$ ellipse that we use to measure the corresponding R$_{\text{out}}$. The errors of vel$_{05}$ were obtained by averaging the vel$_{02}$ and vel$_{10}$ maps and calculating the difference with vel$_{05}$, and for vel$_{95}$, we did the same but using vel$_{98}$ and vel$_{90}$.  These percentiles are representative of the least and most conservative outflow velocities, respectively. All the resulting velocities are listed in Table~\ref{tab:out_prop}.

Then, using the outflow mass, radius, and velocity, and assuming a constant average volume density, as in \citet{Fiore17}, the outflow mass rate can be calculated as:
\begin{equation}
\centering
\dot{\text{M}}_{\rm out} = 3\times {\text v}_{\text{out}} \times \frac{ \text{M}_{\text{out}}}{\text{R}_{\text{out}}} 
\label{eq:massrate}
 \end{equation}
 and the kinetic power as:
 \begin{equation}
\centering
\dot{\text{E}}_{\text{kin}} = \frac{ \dot{\text{M}}_{\rm out}}{2} \,({\text v}_{\text{out}}^2 + 3\sigma)  
 \end{equation}
where $\sigma\sim$FWHM/2.355, with FWHM=W80/1.09, and W$_{80}$ is averaged within the red ellipses shown in Fig.~\ref{fig:out_r}.
In the cases of J1356 and J1430, for which we detect both sides of the outflow, $\dot{\text{M}}_{\rm out}$ and $\dot{\text{E}}_{\text{kin}}$ correspond to the sum of the individual values measured for the blueshifted and redshifted counterparts.

 In Table~\ref{tab:out_prop} we show all the outflow properties described in this Section (i.e., R$_{\text{out}}$, PA, v$_{\text{out}}$, M$_{\rm out}$, $\dot{\text{{M}}}_{\text{{out}}}$, and $\dot{\text{{E}}}_{\text{{kin}}}$), based on the two electron density measurements presented above. Using the [S~II]-based n$_e$ we obtain $\dot{\text{{M}}}_{\text{{out}}} =  3.3 - 6.5$ M$_\odot$yr$^{-1}$ and $\dot{\text{{E}}}_{\text{{kin}}}= 10^{41.1 - 42.0}$ erg s$^{-1}$, whereas using the transauroral-based densities, they are a factor of 3-4 lower: $\dot{\text{{M}}}_{\text{{out}}}$=0.7- 1.6 M$_{\odot}$yr$^{-1}$ and $\dot{\text{{E}}}_{\text{{kin}}}= 10^{40.6 - 41.0}$ erg s$^{-1}$. We also list the corresponding coupling efficiencies in Table \ref{tab:out_prop}, $\xi$= $\dot{\text{{E}}}_{\text{{kin}}}/\text{L}_{\text{bol}}$, which are 0.001-0.01\% if we use the [S~II]-based densities, and 0.0003-0.001\% if we use the trans-auroral-based densities.

\section{Discussion}  \label{discussion}

In Section \ref{results} we presented the results from our analysis of the kinematics of the nuclear (i.e., the central $\sim$2.2 kpc of the QSO2s) and extended 
ionized gas emission, traced by [O~III], of the five QSO2s observed with GTC/MEGARA. From this analysis we concluded that four of the five QSO2s show spatially resolved   [O~III] kinematics and we characterized their outflow properties. Hereafter we discuss the mechanisms that are either responsible or contribute to driving them, their impact on the host galaxies, and the interplay between the ionized and cold molecular gas phases.

\subsection{The role of radio jets in driving the outflows} \label{radio}

\begin{figure*}
\centering
{\par\includegraphics[width=0.49\textwidth]{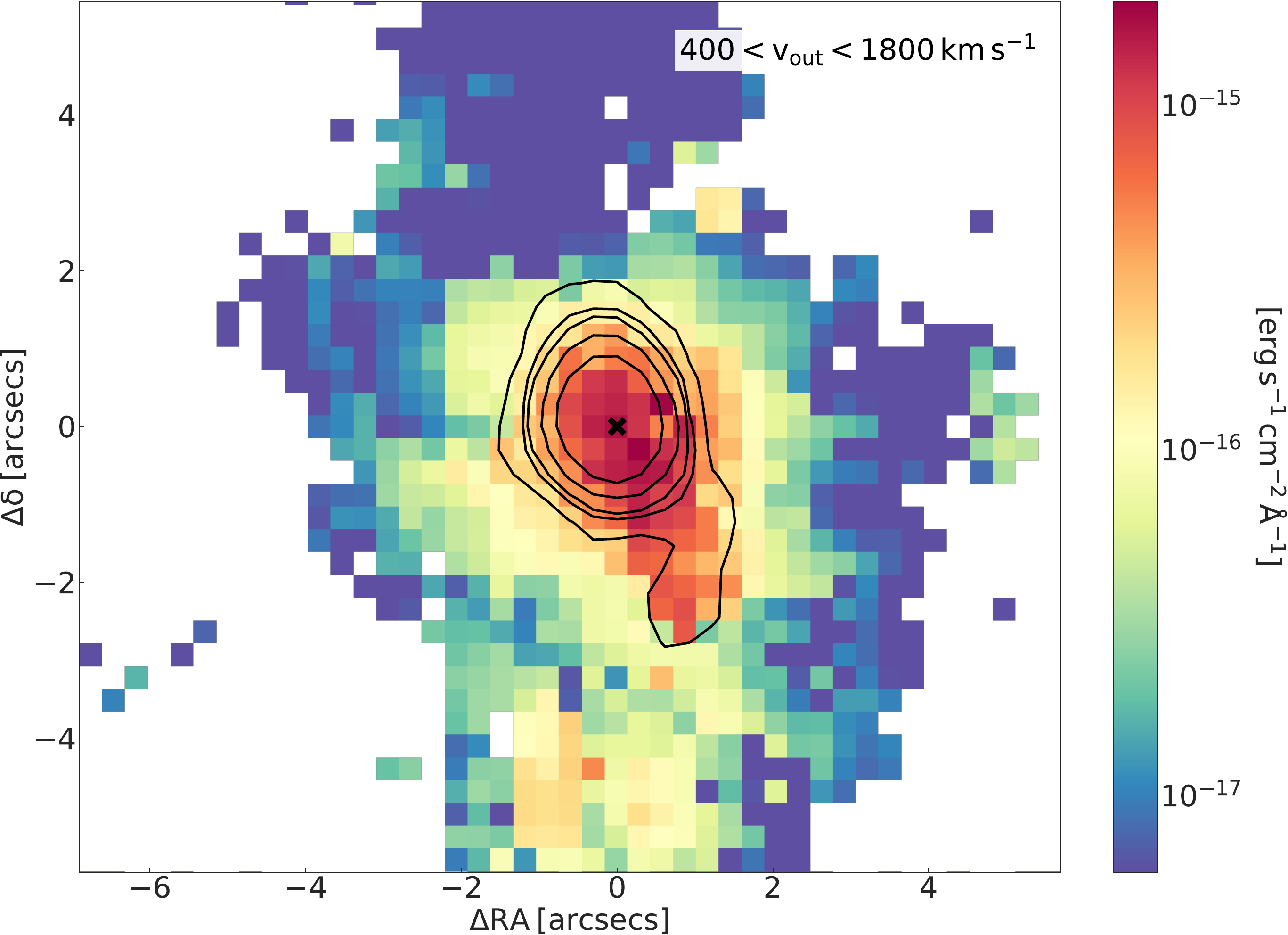}
\includegraphics[width=0.49\textwidth]{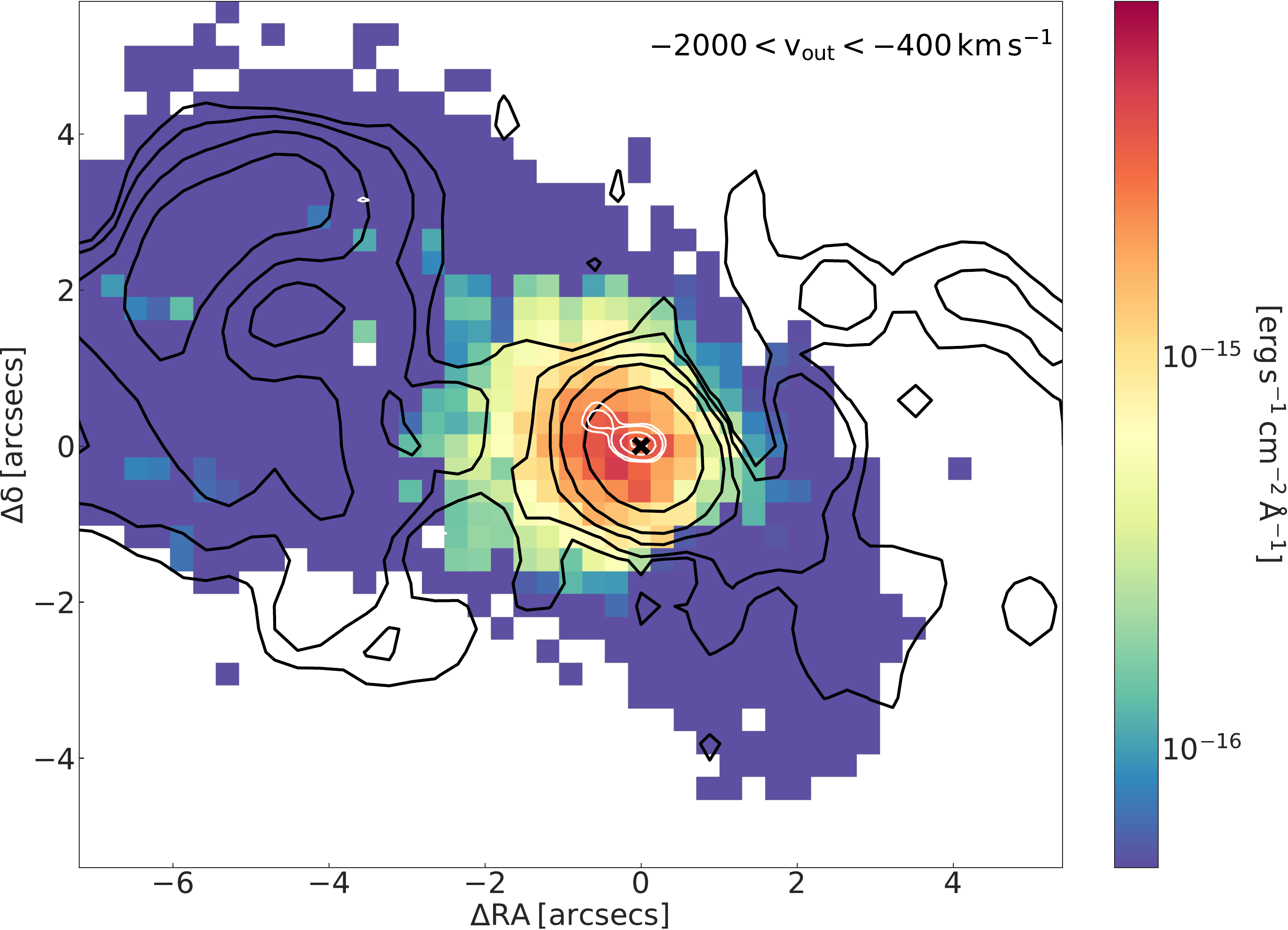}\par
}

\caption {Continuum subtracted [O~III] flux maps of J1356 and J1430 extracted from the cubes in different velocity intervals, indicated in the top right corner of each panel. 
The black cross corresponds to the [O~III] peak, the black contours to the 6 GHz VLA contours at $\sim$1\arcsec~resolution (at 3,10,15,30, and 60$\sigma$ for J1356 and at 3,5,10,20, and 60$\sigma$ for J1430), and the white contours to the 6 GHz VLA contours at $\sim$0.25\arcsec~resolution (at 3,5,15,30, and 60$\sigma$).}
\label{fig:J1356_J1430_radio}
\end{figure*}


As described in Section \ref{sample}, the five QSO2s are radio-quiet and their AGN luminosities are relatively high (10$^{45.5-46}$ erg s$^{-1}$). Considering this, the ionized gas outflows that we studied here should in principle be driven by radiation pressure, generally referred to as ``quasar-mode'' feedback (\citealt{Fabian12}). However, it has been extensively claimed in the literature that low-power compact jets in radio-quiet AGN can compress and shock the surrounding gas as they expand, contributing to drive both ionized and molecular outflows (see e.g., \citealt{Venturi18,Jarvis19,Speranza22,Audibert23}). Alternatively, it has been suggested that the extended radio structures often seen in radio-quiet AGN might be a consequence of AGN winds impacting a higher-density ISM and producing shocks \citep{2019ApJ...887..200F,2023arXiv230615047F}.

In order to explore possible jet-gas interactions in our sample, we compared our outflow flux maps with VLA data at 6 GHz of similar angular resolution. All the targets, except J1509, have observations available at high ($\sim$0.25\arcsec~beam) and low-angular resolution ($\sim$1\arcsec~beam) from \citet{Jarvis19}. These authors reported that in the cases of J1010, J1100, J1356, and J1430, only between 3\% and 6\% of their radio emission can be accounted for by star formation.  The VLA data of J1010 and J1100, at both high and low angular resolutions, do not reveal any extended component. For J1010, 
\citet{Jarvis19} reported a deconvolved size of 0.11\arcsec~($\sim$200 pc), but for J1100 they did not do the same because they relied on the detection of an extended structure only seen in e-MERLIN data. In Figs.~\ref{fig:J1010_extended} and \ref{fig:J1100_extended} we show the low-angular resolution VLA contours of J1010 and J1100.
On the other hand, J1356 and J1430 show extended radio emission    in the VLA data that \citet{Jarvis19} identified with a jet/lobe in the case of J1430, and inconclusive in the case of J1356. For these two galaxies, they reported that the extended radio structures were associated with morphologically and kinematically distinct features detected in [O~III],   indicative of interaction between them. Here we take advantage of the larger FOV covered by GTC/MEGARA to further investigate this.


J1356 shows irregular and messy gas kinematics, as it can be seen from Fig. \ref{fig:J1356_extended} in Appendix \ref{AppendixA}. This is because this QSO2 is hosted in an ongoing merger between two galaxies, the North and South components (see Section 5.1.3 in \citealt{Ramos22}). 
We detect a blueshifted component of the outflow expanding towards the south up to $\sim$ 12.6 kpc from the nucleus. Indeed, this is the maximum radius that we can measure from our data, as we are limited by the FOV (see Fig. \ref{fig:out_r}). This large-scale blueshifted high-velocity [O~III] gas was first reported by \citet{Greene12} and referred to as ``the bubble''. These authors reported a total size of $\sim$20 kpc for this outflowing gas. Here we measure a redshifted counterpart of the outflow expanding towards the south-west (PA$\sim$200\degr) with a maximum extension of $\sim$ 6.8 kpc (see Fig. \ref{fig:out_r} and Section~\ref{J1356} for more details). This redshifted emission, with velocities between 400 and 1800 \kms~relative to the corresponding velocity peak, is perfectly aligned with the 6 GHz radio contours at 1\arcsec~resolution, as can be seen from the left panel of Fig. \ref{fig:J1356_J1430_radio}.  
The radio contours, which have the same angular resolution as our GTC/MEGARA data ($\sim$1\arcsec) extend to the south-west to then bend towards the south, up to -2.5\arcsec~($\sim$5.5 kpc) from the [O~III] emission peak (the black cross in Fig. \ref{fig:J1356_J1430_radio}). At the same distance from the center we observe a sharp decrease of the outflowing flux, of one order of magnitude (from $10^{-15}$ to $10^{-16}$ erg s$^{-1}$ cm$^{-2}\AA^{-1}$). The same contours are superimposed to the vel$_{05}$, vel$_{95}$, and W$_{80}$ maps in  Fig.~\ref{fig:J1356_extended}. The highest outflow velocities and W$_{80}$ are measured within the radio contours and/or immediately next to them,  particularly in the case of the redshifted counterpart of the outflow. 
We argue that this might be observational evidence for a jet-like structure accelerating the ionized gas as it progresses through the galaxy's ISM (see \citealt{Audibert23} and references therein).   However, it is worth noting that such extended radio structure lacks a high resolution counterpart (i.e., in the $\sim$0.2\arcsec~VLA data). This is the reason behind the inconclusive classification of this feature reported by \citet{Jarvis19}, although they favoured the jet/lobe scenario. Another possibility would be a shocked-wind origin for this extended radio structure. 
 Radiatively driven winds could be generating synchrotron emitting shocks in the ISM (e.g., \citealt{Zakamska16,Hwang18,2023arXiv230615047F}). This is  plausible for radio-quiet AGN with bolometric luminosities comparable to J1356 (i.e., L$_{\text{bol}}\approx 10^{45}$ erg s$^{-1}$; \citealt{Faucher12}). 

Focusing on J1430, our kinematic maps reveal an extended [O~III] outflow, with blueshifted velocities increasing to the north-east (R$_{\rm out}$= 3.7 kpc and PA=65\degr) and redshifted velocities towards the south-west (R$_{\rm out}$= 3.1 kpc and PA=198\degr; see middle panels of Fig.~\ref{fig:J1430_map}). The W$_{80}$ map shown in the bottom left panel of the same Figure shows maximum values of $\sim$900\kms~in the same region where the maximum blueshifted velocities are detected. 
The high-angular resolution VLA observations at 6 GHz reveal a compact jet-like structure with a total extent of $\sim$0.8 kpc (PA$\sim$60\degr; \citealt{Harrison15,Jarvis19}; see right panel of Fig. \ref{fig:J1356_J1430_radio}). 
Here we find that the PA of the blueshifted outflow emission coincides with the jet direction (see white contours in the middle left panel of Fig.~\ref{fig:J1430_map}),  in agreement with the results from VLT/MUSE data presented in \citet{Venturi23}. In a recent work, \citet{Audibert23} showed that this compact radio jet, which subtends a small angle with the disc of cold molecular gas, is compressing and accelerating the molecular gas, and driving a lateral outflow that shows enhanced velocity dispersion and higher gas excitation. 
Our data also shows enhanced velocity dispersion (W$_{80}$) in the direction perpendicular to the jet,  reported by \citet{Venturi23}, with maximum values at the jet's forefront (see the bottom left panel of Fig.~\ref{fig:J1430_map}), in agreement with the results obtained for the molecular gas  in this object (\citealt{Audibert23}), and those found for the warm ionised outflows in some nearby Seyfert galaxies (\citealt{Venturi21}).
 This type of jet-gas interaction is typical of the so-called frustrated-jets, i.e., low-power jets that make slow progress through the ISM, disrupting and causing greater damage than powerful collimated jets (\citealt{Wagner11, Nyland18}). 
Hydrodynamical simulations show that clouds of material with densities of n$_e = 300\,\text{cm}^{-3}$ can efficiently confine jets   with P$_{\text{jet}}\leq10^{44}$ erg s$^{-1}$ (\citealt{Mukherjee16, Bicknell18, Talbot22}). 
In the case of J1430, we measure outflow densities  >1700 cm$^{-3}$ using the trans-auroral lines, and \citet{Audibert23} reported a jet power of P$_{\text{jet}}\sim10^{43}$ erg s$^{-1}$. For J1356 we measure outflow densities  >1600 cm$^{-3}$ using the same method, and if we consider that the extended radio structure detected in the low-resolution VLA data is a jet, from these data we  measure P$_{\text{jet}}\sim$10$^{44}$ erg~s$^{-1}$, following the same prescription as in \citet{Audibert23}. Therefore, J1430, and possibly J1356 as well if the extended VLA feature is confirmed as a jet, could be examples of radio-quiet quasars with a low-power jet slowly advancing through the galaxy ISM, shocking and accelerating the surrounding gas, hence contributing to drive the ionized and molecular outflows. A jet-driven outflow is consistent with the kinetic powers that we measure for both QSO2, of $\dot{\text{E}}_{\text{kin}} \sim 10^{41}$ erg s$^{-1}$, two/three orders of magnitude lower than the jet power. 

In the middle and bottom left panels of Fig.~\ref{fig:J1430_map} we show the contours of the brightness temperature ratio (T$_{32}$/T$_{21}$) derived from the CO(3-2) and CO(2-1) lines by \citet{Audibert23} for J1430. The highest values are distributed in the perpendicular direction to the radio jet, slightly offset from the maximum velocities of the CO velocity dispersion (see Figure A.1 in \citealt{Audibert23}). We note that the T$_{32}$/T$_{21}$ contours extend up to where the ionized gas shows the highest outflow velocities and W$_{80}$ values, i.e., at $\sim -0.5 \arcsec$ in the x-axis. This might suggest that the ionized outflow is being accelerated by the combined action of the compact radio jet and the molecular gas that it is entraining. 
On larger scales, we report an enhancement of vel$_{95}$ and W$_{80}$ that overlaps with the extended radio emission traced by the low-resolution 6 GHz data to the south-west (see middle right and bottom left panels of Fig.~\ref{fig:J1430_map}, where the 1\arcsec~resolution radio contours are shown in cyan). In this region the [O~III] emission reaches the highest positive outflow velocities, between 600 and 800 \kms, with corresponding W$_{80}$ values of 600-800 \kms.   

The results that we report here for J1430,   and tentatively for J1356 as well, contribute to strengthen the idea that, even in radio-quiet sources, low-power jets can successfully transfer mechanical energy to the ISM, contributing to drive and/or accelerate ionized gas outflows (e.g. \citealt{Cresci23}). Therefore, radiation-driven winds might not be the only channel contributing to launch kpc-scale outflows in luminous AGN such as our QSO2s, but also low-power compact jets. 



\subsection{The impact of electron density on scaling relations} \label{fiore}

\citet{Harrison14} performed a non-parametric analysis of the [O~III] kinematics for a sample of 16 radio-quiet QSO2s that included J1010, J1100, J1356, and J1430. They used GMOS IFU data, which has a 5$\arcsec \times 3.5\arcsec$ FOV, smaller than MEGARA's (12.5$\arcsec \times 11.3\arcsec$), and assumed the same spherical/multi-cone outflow geometry that we are using here to calculate the mass outflow rates (i.e., Eq.~\ref{eq:massrate}). They measured the electron densities from the [S~II] doublet detected in the SDSS spectra of the QSO2s, as we also do here, and they found values between 200 and 1000 cm$^{-3}$, fully consistent with ours. However, \citet{Harrison14} assumed a value of n$_e$=100 cm$^{-3}$ for all the targets, and the total, extinction-uncorrected H$\beta$ luminosities to work out the outflow masses. For this reason they reported values of M$_{\rm out}$=(2-40)$\times$10$^7$ M$_{\sun}$, which are one order of magnitude larger than ours, of (4-35)$\times$10$^6$ M$_{\sun}$, when we use the [S~II]-based n$_e$. 
Assuming $\sigma$=W$_{\rm 80}$/2.355 and R$_{\rm out}$=R$_{\rm [OIII]}$, they reported outflow mass rates between 3 and 70 $\text{M}_{\odot}\text{yr}^{-1}$ for their sample. For our QSO2s, we measured outflow mass rates between  3.3 and 6.5 $\text{M}_{\odot}\text{yr}^{-1}$ using the [S~II]-based n$_e$ (see Table \ref{tab:out_prop}). If we use the trans-auroral-based n$_e$ values (1600-9800 cm$^{-3}$), reported in the same Table, we measure outflow mass rates between 0.7 and  1.6 $\text{M}_{\odot}\text{yr}^{-1}$.

 A mass outflow rate of $\leq$0.46 M$_{\odot}\text{yr}^{-1}$ was reported by \citet{Ramos19} for J1509 based on near-infrared data from GTC/EMIR. In fact, the outflow rate is an upper limit because the electron density was derived from the total fluxes of the trans-auroral lines, as we are doing here. For J1509 here we measure 1.1$\pm^{0.3}_{0.4}$ M$_{\odot}\text{yr}^{-1}$. For J1430, \citet{Venturi23} reported an ionized mass outflow rate of 40-130 M$_{\odot}$ yr$^{-1}$, measured from VLT/MUSE data on the second radial bin they considered (i.e., between 1 and 2 kpc). The discrepancy between our ionized mass outflow rate and that reported by \citet{Venturi23} mainly comes from 1) the lower values of the electron density employed in the latter work (average value of $\sim$500 cm$^{-3}$ in the second radial bin), 
which were used to calculate the outflow mass on a spaxel-by-spaxel basis; and specially 2) the different methods adopted to estimate the outflow rate. In \citet{Venturi23}, the authors integrated the flux of the broad components fitted to the line profiles in each spaxel, which is significantly higher than the outflow flux estimated from the non-parametric method we used here (see \citealt{Hervella23}), 
and they computed the outflow rate on each spaxel by using $\Delta \rm R_{\rm out}$ equal to the size of each spaxel (0.2\arcsec). Then, they produced radial profiles of $\rm \dot M_{\rm out}$ by summing, in each radial bin, the values measured on each spaxel and then re-scaling using $\Delta \rm R_{\rm out}$/$\Delta \rm R_{\rm bin}$, with $\Delta \rm R_{\rm bin}$=1 kpc. In addition, we note that \citet{Venturi23} reported a decrease in $\rm \dot{M}_{\rm out}$ by a factor of 3--20 if they use v$_{\text{out}  }=\text{v}_{\text{s}}$ instead of v$_{\text{out}  }= \text{v}_{\text{s}}+ \text{FWHM/2}$, where $\text{v}_{\text{s}}$ is the velocity shift of the broad components from the systemic velocity. 


\begin{figure}
\centering
\includegraphics[width=0.50\textwidth]{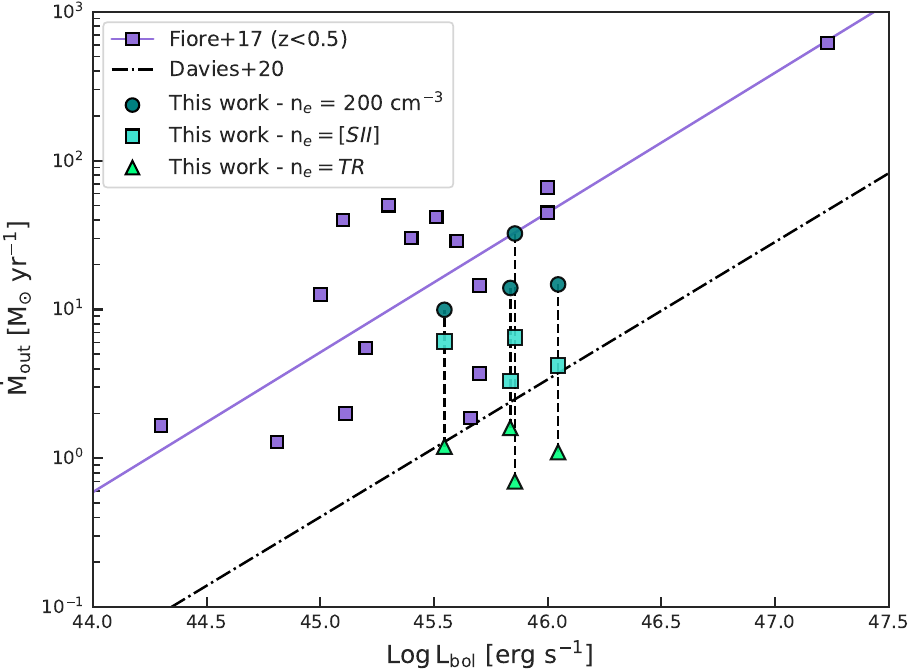}
\caption {Mass outflow rate versus AGN luminosity. The purple line corresponds to the linear fit of the ionized outflow mass rates calculated by \citet{Fiore17} for a sample of AGN at z<0.5 (purple squares) assuming a value of the density of 200 $\text{cm}^{-3}$. The black dot-dashed line corresponds to the linear fit from \citet{Davies20}, that includes the mass rates of \citet{Fiore17}, but corrected by using individual electron densities. The other symbols correspond to the four QSO2s with spatially resolved [O~III] outflows, with different shapes and colours indicating the values of n$_e$ adopted to perform the outflow mass rate calculations: green circles correspond to densities of 200 $\text{cm}^{-3}$, and cyan squares and light-green triangles to those measured from the [S~II] doublet and the trans-auroral lines, respectively. The outflow mass rates of the QSO2s vary up to 2 orders of magnitudes depending on how the n$_e$ is measured.}
\label{fig:fiore_dens}
\end{figure}


To put our results in a broader context, we compare them with those reported by \citet{Fiore17}, who compiled and put together several ionized outflow measurements from the literature, establishing a scaling relation between $\dot{\text{M}}_{\rm out}$ and the AGN luminosity. This comparison is shown in Fig.~\ref{fig:fiore_dens}, where we also represent how our results vary by using different values of n$_e$. When we adopt n$_e = 200 \,\text{cm}^{-3}$, as done by \citet{Fiore17}, we obtain the highest mass outflow rates, of  10-32 $\text{M}_{\odot}\text{yr}^{-1}$ (green circles in Fig.~\ref{fig:fiore_dens}). These values are consistent with the observational scaling relation, defined by the fit of AGN at z<0.5 (i.e., the purple line in the same Figure). However, when the electron densities are calculated individually for each source, the mass outflow rates decrease significantly. Using the n$_e$ measured from the [S~II] doublet, our QSO2s (the cyan squares in Fig.~\ref{fig:fiore_dens}) fall well below the scaling relation. This offset becomes of one order of magnitude when we use the densities derived from the trans-auroral method (see green triangles in Fig. \ref{fig:fiore_dens}),  a trend also shown   in the study conducted by \citet{Santoro20}.
These results highlight the importance of having reliable estimates
of the outflow densities to derive mass outflow rates \citep{2018NatAs...2..198H,Rose18,Davies20, Revalski22, 2023MNRAS.tmp.1680H}, and also, of revisiting the widely used empirical relations. These are likely an upper limit to the real, and much more scattered relation between the mass outflow rate and AGN luminosity (\citealt{Ramos22,Lamperti22}). 

In line with the latter, \citet{Baron19} did not find a linear correlation between the outflow mass rate and AGN luminosity using a  sample of 234 local type-2 AGN with L$_{\text{bol}}=10^{43.3 - 45.8}$ erg s$^{-1}$. They estimated individual electron densities using optical line ratios at the location of the wind (see also \citealt{Revalski18, Revalski18b}), and they reported $\dot{\text{M}}_{\rm out}$ of 1-2 orders of magnitude lower than previous estimates.  \citet{Davies20} also reported lower mass outflow rates than those predicted from the \citet{Fiore17} relation for a sample of 11 Seyfert 2 galaxies using individual n$_e$ values determined as in \citet{Baron19}. By putting together their outflow mass rates, those from \citet{Baron19}, and those from \citet{Fiore17} after calculating individual n$_e$ values for the latter, they proposed a modified version of the \citet{Fiore17} scaling relation, shown as a dot-dashed line in Fig. \ref{fig:fiore_dens}.
They concluded that, even with a larger scatter, a linear correlation is still observed  across 5 orders of magnitude in L$_\text{bol}$, but with outflow mass rates that are lower by about a factor 3, due to the higher n$_e$ values. Our mass rates, both computed using [S~II] and trans-auroral densities, are in agreement with this corrected scaling relation from \citet{Davies20}, which is also shown in Fig.~\ref{fig:fiore_dens}.   
As mentioned in Section~\ref{properties}, here we used two different methods to estimate the outflow densities, but they correspond to total values, as one single Gaussian component was fitted to the corresponding emission lines measured within the SDSS fiber (1.5\arcsec~radius). In \citet{Speranza22} we were able to fit the [S~II] and the trans-auroral lines with a narrow and a broad component for the QSO2 J0945, and we measured a significantly larger n$_e$ value for the latter, being the total value more similar to that of the narrow component (see Table 2 and Fig. 5 in \citealt{Speranza22}). Therefore, it is likely that our QSO2s outflow densities will be even higher than the total n$_{\rm e}$ values employed here, which implies that the outflow masses and mass rates will be even lower.

 Finally, it is worth mentioning that here we are focusing on the influence of the outflow density on the mass outflow rates, but there are other parameters, such as the outflow radius, velocity, and assumed geometry, that strongly impact the outflow rates. Furthermore, the bolometric luminosities also have large uncertainties, of up to $\sim$1 dex, because of the systematic uncertainty on the bolometric correction \citep{Jarvis19}. Nevertheless, \citet{2023A&A...669L...5R} reported good agreement between the AGN luminosities derived from extinction-corrected [O~III] luminosities and hard X-rays for the QSO2 Mrk~477 ($\sim$0.2 dex).

\subsection{A multi-phase view of outflows} \label{SFR}
\begin{figure}
\centering
\includegraphics[width=0.50\textwidth]{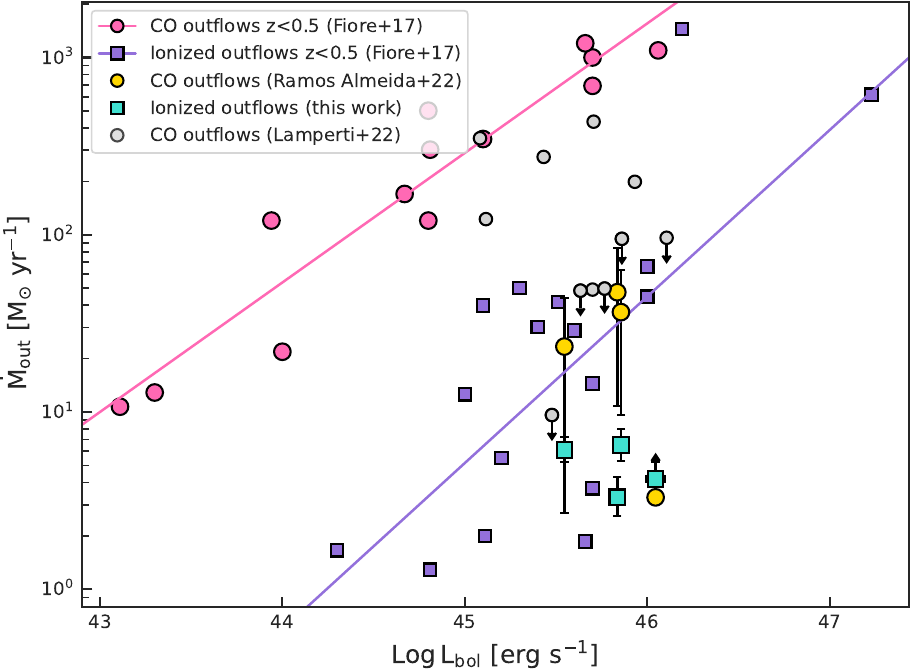}
\caption {Ionized and cold molecular gas mass outflow rates versus AGN  luminosity. The purple and cyan squares are the same as in Fig.~\ref{fig:fiore_dens}, and the pink circles are the cold molecular outflow measurements from \citet{Fiore17}. The corresponding linear fits are indicated with the same colors. 
The yellow circles are the cold molecular mass outflow rates reported by \citet{Ramos22} for the same QSO2s analyzed in this work,  and the grey circles AGN-dominated ULIRGs from \citet{Lamperti22}. 
}
\label{fig:fiore_multi}
\end{figure}

The ionized gas emission corresponds to the warm phase of gas, having typical temperatures of T$\sim$10$^4$ K. As we have shown in Section \ref{fiore}, QSO2s typically have total ionized gas masses of (2-40)$\times$10$^7$ M$_{\sun}$ (\citealt{Harrison14}). These gas masses are much lower than the molecular gas masses measured from CO lines detected in the millimiter range, which range between $\sim$10$^8$ and 10$^{10}$ M$_{\sun}$ for QSO2s at z<0.2 (\citealt{2020MNRAS.498.1560J,Ramos22,2023MNRAS.tmp.3019M}). It has been observationally demonstrated that molecular outflows carry more mass than their ionized counterparts, at least in the local universe, and at AGN luminosities below 10$^{47}$ erg~s$^{-1}$ \citep{Fiore17,Fluetsch21}. However, as it happens with the ionized outflow mass rates, the values used to build the molecular gas  counterpart of the scaling relations (\citealt{2014A&A...562A..21C,Fiore17}) are higher than those found for the general population of quasars \citep{Ramos22},  ULIRGs \citep{Lamperti22}, and Seyfert galaxies \citep{2019A&A...628A..65A,2020arXiv200305663D,Zanchettin21}.

In this work, we aim at providing a more comprehensive view of quasar-driven outflows by putting together the  cold molecular and ionized outflow measurements available for our QSO2s. Using the CO(2-1) transition observed with ALMA at 0.2\arcsec~resolution, \citet{Ramos22} detected spatially resolved cold molecular outflows for the same four QSO2s where we detect spatially resolved ionized outflows (i.e., J1100, J1356, J1430, and J1509). They report higher mass outflow rates, of $\dot{\text{M}}^{\rm H_2}_{\rm out} = 8-16\,\text{M}_{\odot}\text{yr}^{-1}$, than our ionized outflow mass rates, which are $\dot{\text{M}}_{\rm out} =  3.3-6.5\,\text{M}_{\odot}\text{yr}^{-1}$ if we use the [S~II]-based n$_e$, and 0.7-1.6$\,\text{M}_{\odot}\text{yr}^{-1}$ if use the higher densities estimated from the trans-auroral lines.   It is worth mentioning that this comparison is not straightforward because of 1) the different angular resolution of the observations used (0.2\arcsec~in the case of the ALMA observations and 1.2\arcsec~for the MEGARA observations) and 2) the different methodologies employed to derive the outflow properties. In the case of the ALMA CO(2-1) observations, \citet{Ramos22} performed kinematic modeling of the data using $^{\rm 3D}$BAROLO \citep{barolo15} to disentangle rotation from non-circular motions, and integrated the flux of the high-velocity gas detected along the kinematic minor axis, using an slit-simulated aperture of 0.2\arcsec~width (see \citealt{Ramos22} for further details). Instead, here we integrated all the flux of the high-velocity ionized gas within ellipses defined by the 3$\sigma$ contours. However, despite these differences in angular resolution and methodology, we note that the molecular outflow measurements are rather conservative, and different methodologies will most likely result in even higher molecular outflow masses and/or outflow rates. As an example, \citet{Audibert23} considered four different scenarios, from most to least conservative, to calculate the outflow mass rates of J1430 using the same CO(2-1) observations as \citet{Ramos22}, and reported values from 6.7 to 44 M$_{\sun}$~yr$^{-1}$.

Higher molecular outflow rates are found because, even though the molecular outflow velocities are lower ($\le$300\kms) than the ionized outflow velocities that we measure  here (500$\le|v_{ \rm out}|\le$1300\kms), the cold molecular outflows carry more mass and are more compact. For the latter \citet{Ramos22} reported masses of (1-10)$\times$10$^7$ M$_{\sun}$ and radii $\le$3 kpc, whereas here we measure (1-35)$\times$10$^6$ M$_{\sun}$ and radii of  3.1 - 12.6 kpc for the ionized outflows (see Table \ref{tab:out_prop}).
Thus,  for our sample, the molecular mass outflow rates are on average 2-3 times larger than the ionized mass outflow rates. However, from the comparison between the ionized and cold molecular scaling relations in \citet{Fiore17}, the gap between the two gas phases at the bolometric luminosities of the QSO2s is expected to be almost of two orders of magnitude, that is a factor of $\sim$50. This is shown in Fig.~\ref{fig:fiore_multi}, where the observational scaling relations for the molecular and ionized outflows are the pink and purple lines, respectively. The molecular outflow mass rates measured by \citet{Ramos22} are the yellow circles, and the corresponding ionized mass outflow rates, derived using the [S~II]-based densities, are the cyan squares. If we consider the ionized outflow rates obtained using the trans-auroral-based densities, they are $\sim$12 times lower than the molecular outflow rates.
Using 0.2\arcsec~resolution CO(2-1) observations of a sample of nearby ULIRGs, \citet{Lamperti22} measured molecular outflow rates of 16-145 M$_{\odot}$ yr$^{-1}$ for six AGN-dominated ULIRGs with similar bolometric luminosities as our QSO2s, and upper limits of $\leq 3-32$  M$_{\odot}$ yr$^{-1}$ for another five AGN-dominated ULIRGs  (see grey circles in Fig.~\ref{fig:fiore_multi}).


These results highlight that the $\dot{\text{M}}_{\rm out}$ versus L$_{\text{bol}}$ empirical relation might be just the upper envelope of a much more scattered relation between these two quantities. 
Based on our multi-phase analysis of this small sample of QSO2s, we find that the cold molecular outflows are indeed carrying more mass  and are more compact than their ionized counterparts. This introduces a significant offset in $\dot{\text{M}}_{\rm out}$ between the two gas phases. 
A larger sample with multi-phase outflow measurements in the same targets, which is the ultimate goal of the QSOFEED project \citep{Ramos22}, is needed to derive statistically significant conclusions. If the linear trend is still observed in a larger sample, it might just need to be re-scaled \citep{Davies20}. Otherwise, as discussed in \citet{Ramos22}, outflow mass rates might not depend only on AGN luminosity, but in other factors such as the coupling between the jets and/or the winds and the ionized and molecular gas, and the amount and distribution of gas in the central regions. A combination of these two scenarios is also possible, in which case the scaling relation would persist, albeit with larger scatter.


Using the SFRs of the QSO2s, derived from their far-infrared luminosities (see Table \ref{tab:sample}), we can calculate the mass loading factors ($\eta = \dot{\text{M}}_{\rm out}/$SFR) of the ionized outflows, which are $\eta$=0.1-0.2. These values cover a smaller range than those measured for the cold molecular phase ($\eta_{\rm H_2}$=0.1-1.3; \citealt{Ramos22}). If we sum the loading factors of the two phases, $\eta_{\rm tot} = \eta + \eta_{\rm H_2}$, we get $\eta_{\text{tot}}$= 0.3+1.3= 1.6 for J1430, and $\eta_{\text{tot}}$=0.2-0.5 for the other QSO2s. Values of $\eta_{\text{tot}}\le$1 imply, in principle, that outflows are less effective removing gas than star formation, and also, that star-formation could be potentially driving them. However, a few caveats must be mentioned in this respect. First, we argue that the outflows, both molecular and ionized, are AGN-driven. The high outflow velocities that we measure cannot be accounted for by star formation \citep{Speranza22}. Second, the SFR measurements are calculated from the total far-infrared luminosity, and therefore are galaxy-wide SFRs. The outflows, on the other hand, are more compact, specially the molecular outflows (r$_{\rm out}$=0.5-3 kpc; \citealt{Ramos22}). Indeed, despite the relatively low $\eta_{\rm H_2}$ values of the QSO2s, \citet{Ramos22} and \citet{Audibert23} reported that these outflows, in some cases in combination with compact jets, are modifying the molecular gas distribution in the central kiloparsec of the galaxies. High angular resolution IFU observations are needed to compare the spatial distribution of recent star formation and AGN outflows to really evaluate the impact of the latter. 

Another caveat is the different timescale of the star formation probed by the far-infrared luminosity and the outflows. Considering that the QSO2 outflows have dynamical timescales of $\sim$1-10 Myr \citep{Bessiere22,Ramos22}, their properties should be compared with recent star formation.  This can be probed using resolved stellar population analyses (\citealt{Bessiere22}) or the polycyclic aromatic hydrocarbon (PAH) features detected in the mid-infrared, which are good probes of young and massive stars, but whose properties vary with distance from the AGN (\citealt{2022A&A...666L...5G,2023A&A...669L...5R}). These types of studies can reveal whether recent star formation is prevented or triggered by the outflows, or if both can simultaneously happen (\citealt{Cresci15,Carniani16,Bessiere22}). Indeed, for the five QSO2s studied here, we will study the spatial distribution of the PAH features using recently awarded Cycle 2 James Webb Space Telescope (JWST) observations (PI: Ramos Almeida, proposal 3655), 
and we will compare it with the spatially resolved outflows presented here. From our measurements of the outflow radii (R$_\text{out}$ = 3.3 -12.7 kpc) and velocities (v$_{\text{out}}$ = 500-1300 \kms) we obtain dynamical timescales for the ionized outflows of t$_\text{dyn}$ = R$_\text{out}$/ v$_{\text{out}}$=3-20 Myr, consistent with those of the molecular outflows and the young stellar populations that we will probe with JWST.


 \section{Summary and conclusions}
 \label{conclusion}

We have investigated the [O~III]$\lambda$5007$\AA$ emission of five nearby QSO2s (z$\sim$0.1)  from the QSOFEED sample using seeing-limited, high spectral resolution observations obtained with GTC/MEGARA. The IFU capabilities of this instrument allowed us to explore the nuclear emission from the central 1.2\arcsec~($\sim$2.2 kpc) of the galaxies, and the ionized extended emission across a FOV of 12.5\arcsec$\times$11.3\arcsec~($\sim$
23$\times$21 kpc$^2$). We characterize the [O~III] kinematics of the QSO2s and, thanks to a recent high angular resolution study of their CO(2-1) emission (\citealt{Ramos22}), we provide a comprehensive multi-wavelength view of their gas content and kinematics. The main conclusions from our work are listed as it follows.

\begin{itemize}

\item The nuclear spectra of the 5 QSO2s present signatures of outflowing gas. Broad components with 1300$\leq$FWHM$\leq$2200 \kms~are detected, all with blueshifted velocities ranging from -40 to -430 \kms. In all the QSO2s but in J1509, we also fitted intermediate components to reproduce the line profiles, with 340$\leq$FWHM$\leq$900 \kms, some of them blueshifted (with velocities of up to -365 \kms) and others redshifted (up to 280 \kms) with respect to the systemic velocity.

\item We detect kinematically disturbed gas in the five QSO2s (high velocities and turbulence) that we attribute to outflowing gas. These outflows are spatially resolved in all the QSO2s except in J1010. For the four QSO2s in which the outflows are resolved, we measure radii between  3.1 and  12.6 kpc. In J1356 and J1430 we detect both sides of the outflow, approaching and receding, and in J1100 and J1509, only the approaching side.

\item The receding side of the outflow in J1356, and the approaching side of the one in J1430 are well aligned with the extended radio  structures detected in VLA observations. 
This co-spatiality might be indicative of compact, low-power jets driving or contributing to accelerate the ionized outflows in these radio-quiet QSO2s.  Alternatively, radiativelly-driven winds could be inducing shocks in the ISM and producing the extended structures, specially in the case of J1356, where the nature of the extended radio structure is unclear.

\item The electron densities measured from the [S~II]$\lambda\lambda6716,6713$ doublet range between 300 and 1000 cm$^{-3}$, and the corresponding mass outflow rates are  3.3-6.5 M$_\odot$ yr$^{-1}$. If instead we use the higher electron densities measured from the trans-auroral lines (1600-9800 cm$^{-3}$), the mass outflow rates decrease by a factor of 3-4 (0.7- 1.6 M$_\odot$ yr$^{-1}$). 

\item The ionized and cold molecular mass outflow rates measured for the four QSO2s with spatially resolved outflows are well below the empirical scaling relation from \citet{Fiore17}. The cold molecular outflows carry more mass and are more compact than their ionized counterparts, but the expected gap between the two phases at the bolometric luminosities of the QSO2s decreases from a factor of $\sim$50 to 2-12 depending of the method used to calculate the electron densities. We argue that empirical scaling relations might need to be re-scaled by using precise multi-phase outflow measurements of the same targets.  


\item Using our ionized and cold molecular outflow mass rates and the total SFRs measured from the far-infrared luminosities we find a total mass loading factor $\eta_{\text{tot}} = (\dot{\text{M}}_\text{out} + \dot{\text{M}}_\text{out}^{\rm H_2})/$SFR =  1.6 in J1430 (the Teacup), and  $\eta_{\text{tot}} = 0.2 - 0.5$ in the other three QSO2s. Despite these low values, we argue that the outflows are AGN-driven, based on different considerations (e.g., the high outflow velocities).



\end{itemize}


We do not find a significant impact of quasar-driven outflows on the global star formation rates when considering the energy budget of the molecular and ionized outflows together. However, considering the dynamical timescales of the outflows, of 3-20 Myr in the case of the ionized gas and 1-10 Myr for the molecular gas, spatially resolved measurements of the recent star formation in these targets are needed in order to fairly evaluate this impact. Our forthcoming JWST observations of the five QSO2s studied here will permit us to do so and thus advance in our understanding of AGN feedback in the local universe. 

\begin{acknowledgements}
Based on observations made with the GTC, installed in the Spanish Observatorio del Roque de los Muchachos of the Instituto de Astrofísica de Canarias, in the island of La Palma. The authors are extremely grateful to the GTC staff for their constant and enthusiastic support.  This work has been supported by the EU H2020-MSCA-ITN-2019 Project 860744 “BiD4BESt: Big Data applications for black hole Evolution STudies.”
CRA and GS acknowledge the
project “Feeding and feedback in active galaxies”, with reference PID2019-106027GB-C42, funded by MICINN-AEI/10.13039/501100011033. CRA also acknowledges the project ``Quantifying the
impact of quasar feedback on galaxy evolution'', with reference
EUR2020-112266, funded by MICINN-AEI/10.13039/501100011033 and the European Union NextGenerationEU/PRTR. OGM acknowledge support from the PAPIIT/DGAPA project IN109123 from UNAM and the “Frontera de la Ciencia” project CF-2023-G-100 from CONACyT. AL is partly supported by the PRIN MIUR 2017 prot. 20173ML3WW 002 ‘Opening the ALMA window on the cosmic evolution of gas, stars, and massive black holes’. CRA thanks the Kavli Institute for Cosmology of the University of Cambridge for their hospitality while working on this paper, and the IAC Severo Ochoa Program for the corresponding financial support.  GS and CRA thank Begoña García Lorenzo, Giacomo Venturi, and Chris Harrison for helpful discussions and Patricia Bessiere for sharing the SDSS outflow measurements of the QSOFEED sample. AL thanks Maria Vittoria Zanchettin for helpful discussions. We finally thank the anonymous referee for useful and constructive suggestions.

\end{acknowledgements}

\medskip
\bibliography{./sample}
 
\appendix
\section{Kinematics maps}
\label{AppendixA}

In Section~\ref{extended}  we present and describe the kinematic maps of J1430. Here we do the same for the rest of the sample, comparing our results with previous studies.  The outflow orientations and maximum extensions are measured as described in Section~\ref{properties}. For a more detailed description of the five QSO2s, including their molecular gas content, distribution, and kinematics, we refer the reader to \citet{Ramos22}.

\subsection{J1010}
\label{J1010}

\begin{figure*}[!h]
\centering
\includegraphics[width=1.\textwidth]{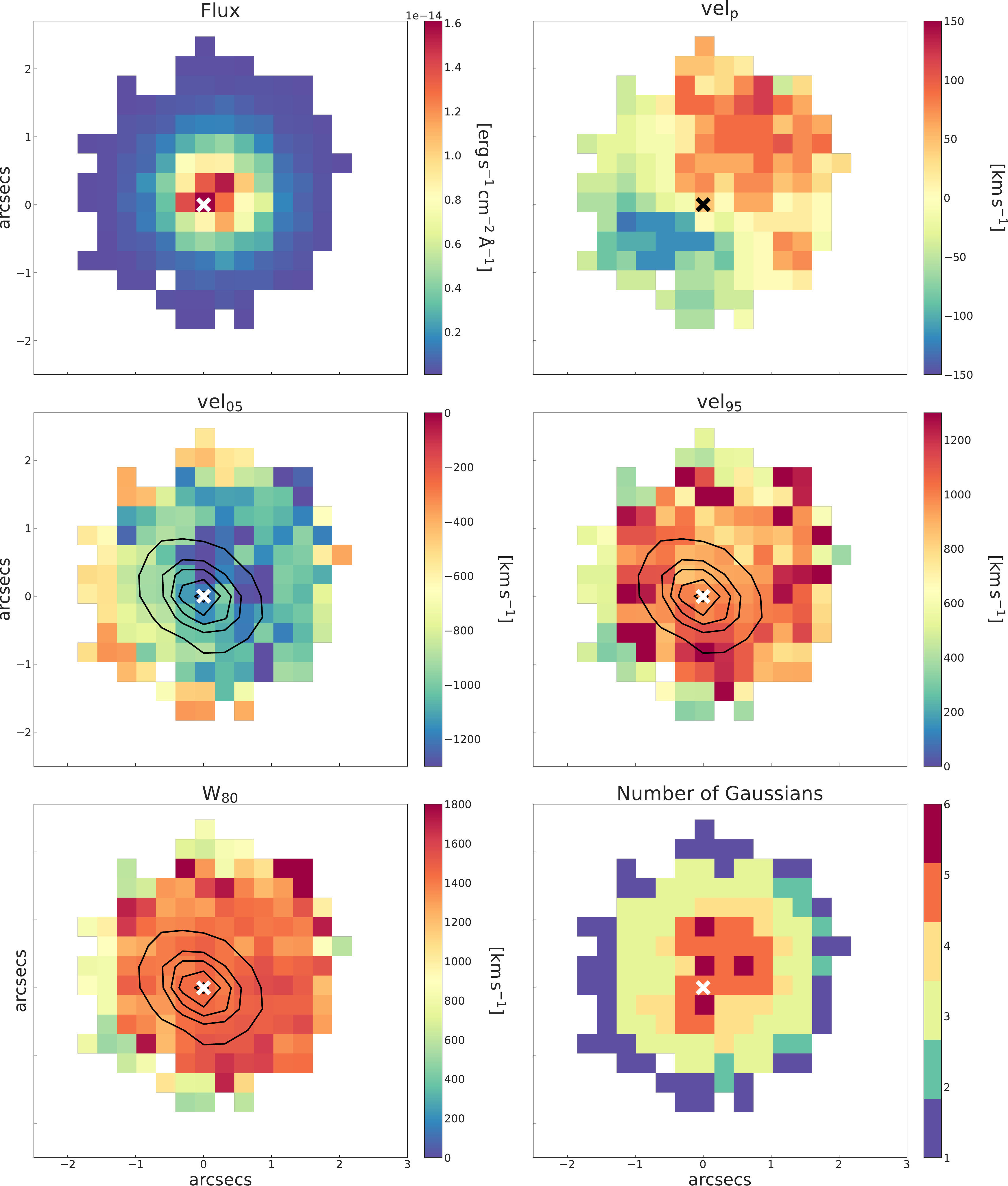}
\caption {Same as in Fig.~\ref{fig:J1430_map}, but for J1010.  The $\sim$1\arcsec~resolution 6 GHz VLA contours at 3,10,15,20, and 25$\sigma$ from  \citet{Jarvis19} are superimposed to the vel$_{05}$, vel$_{95}$, and W$_{80}$ maps.} 
\label{fig:J1010_extended}
\end{figure*}

This QSO2 is hosted in an early-type galaxy which is interacting with a smaller companion located at 13 kpc to the south-west. The CO(2-1) morphology of the QSO2, studied in \citet{Ramos22} at 0.8\arcsec~resolution, shows two peaks that are 1.25 kpc apart. This peculiar CO morphology is most likely produced by gas removal/excitation in the north-south direction. 
 We show the kinematic maps of J1010 in Fig.~\ref{fig:J1010_extended}.  Its [O~III] emission is round and centrally peaked, having a diameter of $\sim$4\arcsec~(7.2 kpc),    directly estimated from the kinematic maps.  
 The vel$_p$ map shows gas rotation, with negative velocities in the south-east and positive velocities to the north-west. This rotation pattern was first reported by \citet{Harrison14}, who used Gemini/GMOS observations of the central 5$\times$3.5 arcsec$^2$ of the QSO2. The vel$_{05}$ map shows the highest velocities towards the north-west, and the vel$_{95}$ map in the opposite direction. However, for this QSO2 we find that the [O~III] emission is not spatially resolved (see Section~\ref{PSF}), explaining the detection of high values of W$_{80}$ ($\geq$1200 \kms) across almost the entire FOV (see bottom left panel of Fig.~\ref{fig:J1010_extended}). Therefore,   based on our analysis we find that our seeing-limited data does not allow us to resolve the ionized outflow or disentangle its orientation. However, \citet{Ramos22} inspected archival VLT/MUSE data (PI: G. Venturi, 0104.B-0476) and reported higher [O~III] FWHMs in the north-west region of the galaxy. 
 
\subsection{J1100}
\label{J1100}

\begin{figure*}[!h]
\centering
\includegraphics[width=1.\textwidth]{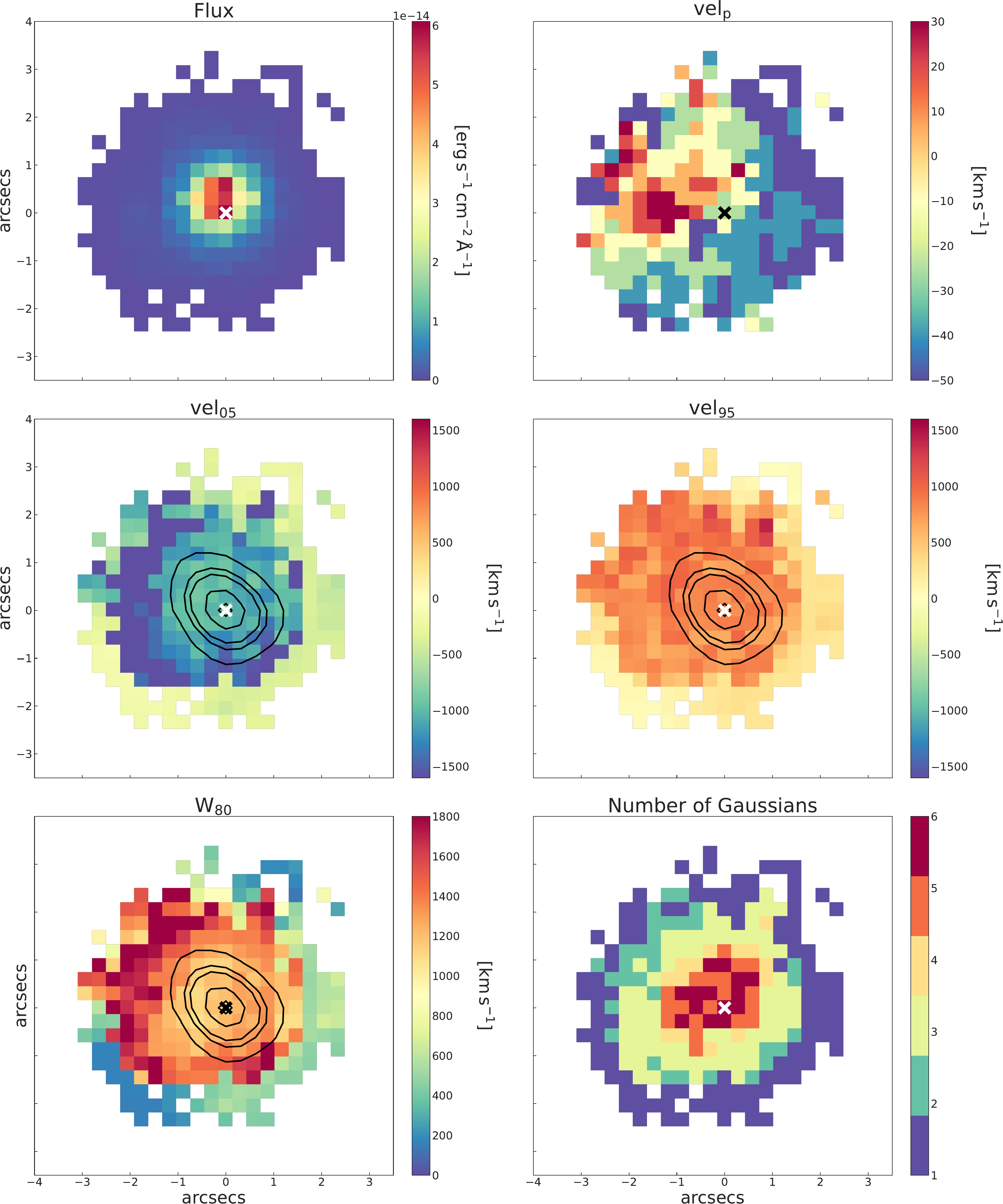}
\caption {Same as in Fig.~\ref{fig:J1430_map}, but for J1100.  The $\sim$1\arcsec~resolution 6 GHz VLA contours at 3,10,15,30, and 40$\sigma$ from \citet{Jarvis19} are superimposed to the vel$_{05}$, vel$_{95}$, and W$_{80}$ maps.}
\label{fig:J1100_extended}
\end{figure*}

J1100 is a barred spiral galaxy, moderately inclined (i=38\degr), the south being the near side \citep{Ramos22}. \citet{Fischer18} presented HST/ACS imaging observations of J1100 that show rounded [O~III] emission, of 1.1$\arcsec$ ($\sim$2 kpc) in radius, with a small tail towards the south-east. Our GTC/MEGARA kinematic maps, shown in Fig.~\ref{fig:J1100_extended}, also show a round [O~III] morphology, with a radius of $\sim$3\arcsec~($\sim$5.5 kpc).

The velocity peak map shows some deviations from regular rotation, also reported from GMOS/IFU data by \citet{Harrison14}, which are most likely due to the stellar bar present in this galaxy. Despite that, we can distinguish positive velocities to be dominant in the north-east region and negative in the south-weast. This rotation pattern coincides with the kinematics of the cold molecular gas studied with ALMA at 0.2\arcsec~resolution \citep{Ramos22}.

\citet{Harrison14} reported broad [O~III] profiles across the entire FOV and velocity blueshifts increasing towards the south-east, but with large uncertainty. According to our analysis, presented in Section \ref{PSF}, the [O~III] emission is spatially resolved, and thus we can clearly identify high blueshifted velocities ($\geq1200$ \kms) on the north-east side of the galaxy that correspond to broad line profiles (W$_{80}\geq 1600$ \kms). Using HST/STIS data of J1100 and a slit orientation of $-19^{\circ}$, \citet{Fischer18} reported the presence of blueshifted velocities to the south-east, with FWHM of 1780 \kms. This is in agreement with our results if we look at the values of the kinematic maps at PA=$-19^{\circ}$, but thanks to the IFU capabilities of MEGARA, we can see from the maps that the preferential orientation of the outflow (i.e., the maximum velocities and FWHMs) is to the north-east, with PA$= 63^{\circ}$ and R$_{\rm out}$=2.8\arcsec~($\sim$5.1 kpc). Considering this new information, for the approaching side of the outflow to be detected in the north, which is the far side of the galaxy, it must subtend a large angle relative to the galaxy disc.

\subsection{J1356}
\label{J1356}

\begin{figure*}[!h]
\centering
\includegraphics[width=1.\textwidth]{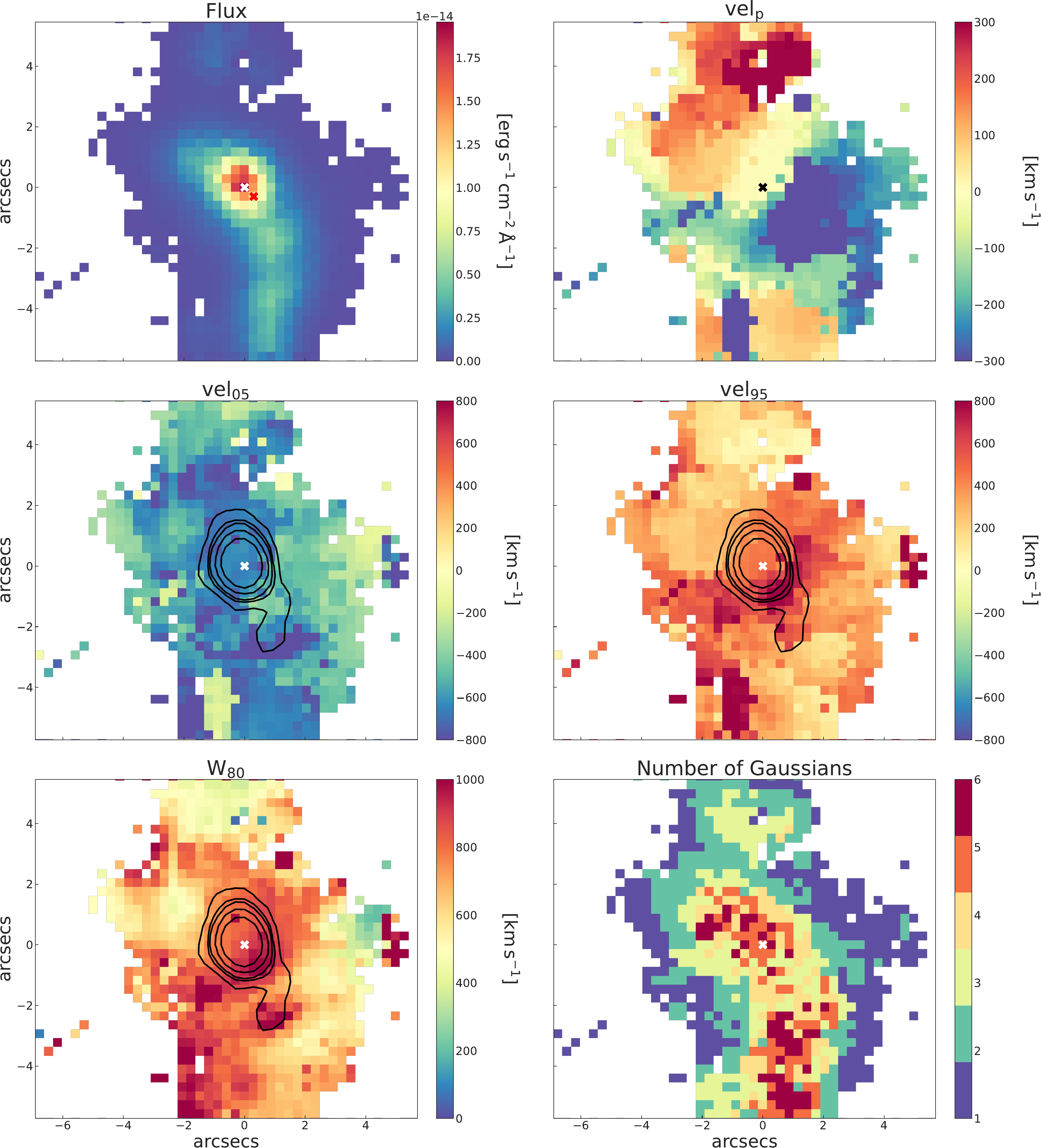}
\caption {Same as in Fig.~\ref{fig:J1430_map}, but for J1356. The red cross in the top left panel marks the AGN position, that is offset by one pixel from the [O~III] peak (indicated by the white cross). The $\sim$1\arcsec~resolution 6 GHz VLA contours at 3,10,15,30, and 60 $\sigma$ from  from \citet{Jarvis19} are superimposed to the vel$_{05}$, vel$_{95}$, and W$_{80}$ maps.} 
\label{fig:J1356_extended}
\end{figure*}

J1356 is a merging system with two nuclei, north and south, observed in the optical (\citealt{Greene12, Harrison14}) and in molecular gas (\citealt{Sun14, Ramos22}). From our GTC/MEGARA data we find the north nucleus to be the brightest, as it is also the case for CO. According to  \citet{Ramos22}, it contains $\sim$55\% of the total molecular mass in the system. The kinematic maps shown in Fig. \ref{fig:J1356_extended} are then centred at the [O~III] peak of the north nucleus.

The ionized gas shows positive velocities north-east of the north galaxy, and negative to the south-west, but the major axis is not aligned with the CO major axis (PA=110$^{\circ}$) reported by \citet{Ramos22}. Overall, the kinematics are complex and disturbed because of the ongoing merger.
 In the case of this QSO2 we detect both the approaching and receding sides of the outflow, with higher values of vel$_{05}$ and vel$_{95}$ corresponding to an enhancement in W$_{80}$. We detect the [O~III] outflowing ``bubble'' proposed by \citet{Greene12}, which is expanding towards the south up to $\sim$12 kpc from the north nucleus (maximum R$_{\rm out}\sim$ 12.6 kpc),  with a PA$\sim197^{\circ}$. In fact, this outflow radius is limited by the size of the GTC/MEGARA FOV, but according to \citet{Greene12}, the size of the bubble is $\sim$20 kpc. The widths measured in this south region are W$_{80}\approx800-1000$\kms. A receding outflow component is also detected to the south with a similar PA ($\sim200^{\circ}$) but less extended, up to  6.8 kpc, and with the bulk of the flux confined within 2.5\arcsec~($\sim$5.5 kpc). We find this redshifted outflow component to be most likely accelerated by extended radio-structure detected in VLA at 1\arcsec~resolution (see left panel of Fig. \ref{fig:J1356_J1430_radio}), as discussed in Section~\ref{radio}. The blueshifted ``bubble'' and the redshifted outflow might correspond to different outflow episodes, considering their different spatial and dynamical timescales (we measure $\sim$ 20 Myr for the blueshifted ``bubble'' and $\sim$15 Myr for the redshifted outflow). \citet{Ramos22} also reported redshifted outflow velocities for the molecular outflow that they detected in this QSO2, albeit much more compact (R$_{\rm out}$=0.4 kpc; t$_{\rm dyn}$=1.4 Myr).

\subsection{J1430}
\label{J1430}

The maps corresponding to J1430 are shown in Fig.~\ref{fig:J1430_map}. This QSO2 has been extensively studied   in different wavelength ranges, and it is known as the Teacup. As mentioned in Section~\ref{extended}, our kinematic maps are similar to those presented in \citet{Harrison14}, where Gemini/GMOS data were used,  in \citealt{Harrison15}, based on VLT/VIMOS data, in \citet{Ramos17}, based on VLT/SINFONI data, and most recently in \citet{Venturi23}, using VLT/MUSE observations. In the vel$_p$ map, we see positive velocities to the north-east and negative to the south-west, having a total velocity gradient of 600 \kms~(see the top right panel of Fig.~\ref{fig:J1430_map}). %
\citet{Ramos17} reported a spatially resolved ionized outflow  with a PA$\sim72-74^{\circ}$, which is comparable to the PA that we measure in this study ($\approx 65^{\circ}$) for the approaching side of the outflow, which exhibits velocities reaching up to -1000 \kms. This is consistent with the blueshifted outflow reported by \citet{Harrison15}, oriented as the central small-scale radio jet. \citet{Venturi23} reported an enhancement of the velocity dispersion perpendicular to this radio jet, also found for the molecular gas (\citealt{Ramos22,Audibert23}). Here we confirm these results and, furthermore, our observations reveal the receding side of the outflow, extending in the opposite direction, with a PA of approximately 200$^{\circ}$. This was first reported by \citet{Keel17}, who measured velocities of up to $\pm$1000 \kms~in the same area where we detect blue and red wings. According to \citet{Ramos22}, considering that the east is the far side of the galaxy, for the ionized outflow to be blueshifted to the east, it must subtend a large angle relative to the galaxy disc. 

The CO(2-1) morphology of J1430 exhibits a double peaked structure perpendicular to the radio jet, as reported by \citet{Ramos22}. Differently, the optical emission displays a single peak, as observed in previous studies by \citet{keel12, Harrison15, Ramos17, Venturi23} and our own work. \citet{Ramos22} suggested that this difference in the distribution of molecular gas compared to ionized gas may be attributed to the combined action of the molecular outflow and the jet. They found the cold molecular outflow to be redshifted to the east and blueshifted to the west, that is the opposite trend compared to  the ionized outflow,  where blueshifted velocities are detected to the nort-east and redshifted to the south-west. However, it noteworthy that \citet{Ramos22} studied the molecular outflow emission along the minor axis only. In the more detailed study of \citet{Audibert23}, the blue side of the outflow would be oriented to the south-east, and the red to the north-west. Therefore, the ionized and cold molecular outflow in J1430 do not share either orientation or PA. 


\subsection{J1509}
\label{J1509}

\begin{figure*}[!h]
\centering
\includegraphics[width=1.\textwidth]{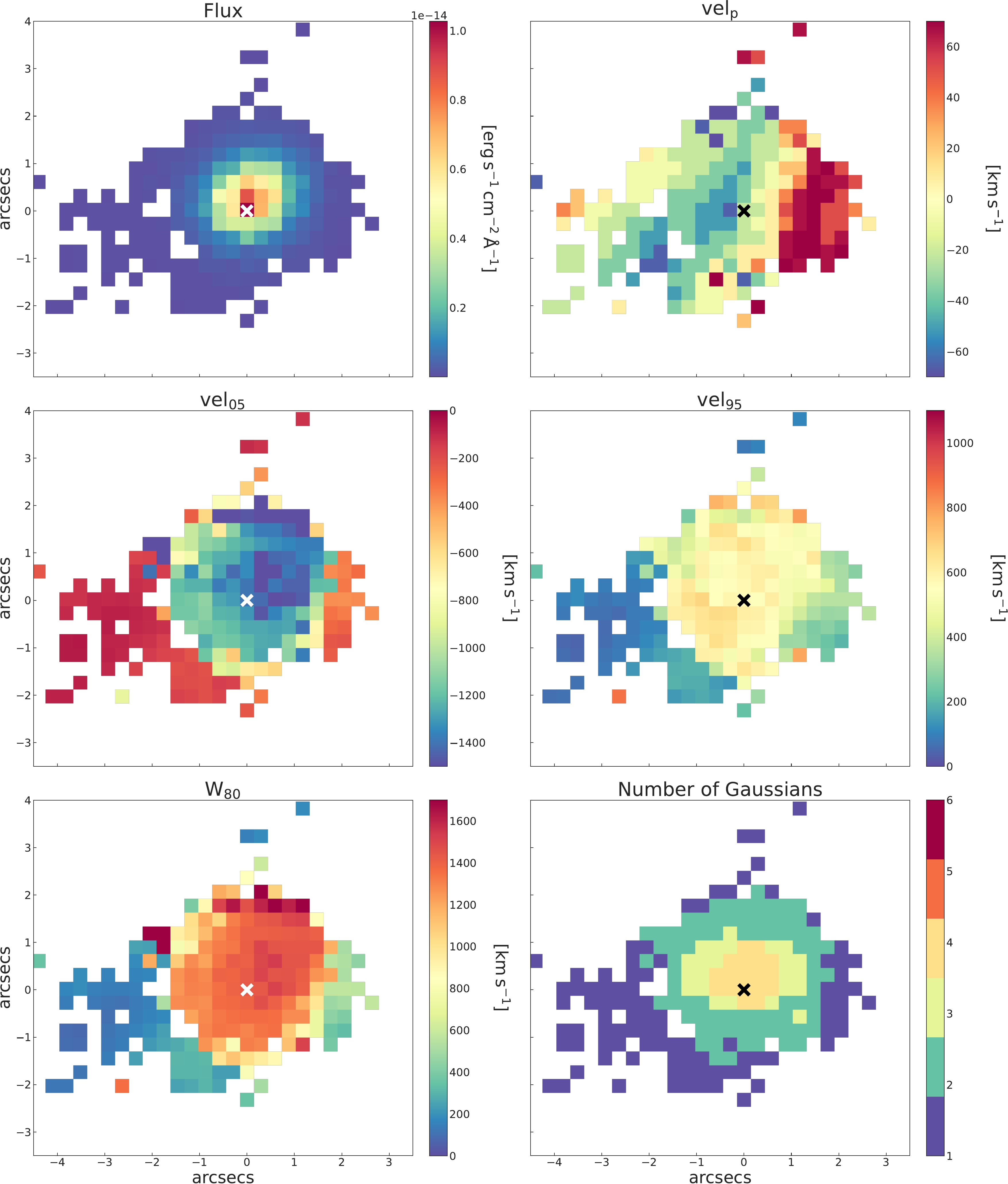}
\caption {Same as in Fig.~\ref{fig:J1430_map}, but for J1509.}
\label{fig:J1509_extended}
\end{figure*}

The [O~III] morphology of J1509, a barred galaxy whose near side is the north,   shows a compact disc of $\sim$1.5\arcsec~(3 kpc) in radius, as directly measured from the MEGARA maps shown in Fig. \ref{fig:J1509_extended}, and a fish tail-like feature to the south-east. This is the only QSO2 in the sample not included in the kinematic study of \citet{Harrison14}  and lacking VLA radio observations. Indeed, to the best of our knowledge, these are the first optical IFU observations of this QSO2. The kinematic maps shown in Fig. \ref{fig:J1509_extended} reveal gas rotation, with negative velocities to the east and positive to the west. This coincides with the kinematics of the molecular gas studied with ALMA by \citet{Ramos22}. 

A multi-phase outflow was reported in this QSO2 by \citet{Ramos19} using near-infrared GTC/EMIR long-slit observations. They detected blueshifted ionized and warm molecular gas, with FWHM$\sim$1500-1700\kms. The slit was placed roughly along the minor axis of the galaxy, PA=-16\degr. Our vel$_{05}$ map, shown in the middle-left panel of Fig. \ref{fig:J1509_extended}, indeed shows maximum blueshifted velocities to the north-west, which is also the region with the highest values of W$_{\rm 80}$=1500-1600\kms. Therefore, we confirm the detection of the approaching side of the outflow to the north-west, with a PA$\sim$40\degr~and R$_{\rm out}$= 3.8 kpc. 
The ALMA CO(2-1) observations also show a very compact and coplanar molecular outflow, blueshifted to the north, and redshifted to the south. 
We do not observe the redshifted side of the ionized outflow, since the vel$_{95}$ map does not show any structure ascribable to an outflow. If present, it must be hidden by the galaxy disc. Thus, for the blueshifted side of the ionized outflow to the detected in the north, which is the near side, the outflow must subtend a small angle with the galaxy disc.

\end{document}